%% file: RAIDS-IEEE-main.tex
\documentclass[conference, 10pt]{IEEEtran}  
\IEEEoverridecommandlockouts
 \pdfoutput=1
\usepackage[space]{cite}
\usepackage{amsmath,amssymb,amsfonts}
\usepackage{algorithm, hyperref, algorithmicx}
\usepackage{algpseudocode}
\usepackage{graphicx}
\usepackage{textcomp}
\usepackage{xcolor}
\usepackage{comment}
\usepackage{tikz, booktabs}
\usepackage{subfigure}
\usepackage{enumitem}
\usepackage{amsmath, tikz}
\usepackage{comment, multirow, enumitem, booktabs, color, yfonts, graphicx, tikz}
\usepackage{algorithmicx, algpseudocode}
\usepackage{algorithm, amsmath, bm}
\usepackage{enumitem}
\usepackage{pifont}
\usepackage{subfigure}
\usepackage[normalem]{ulem}
\usepackage{booktabs}  

\usepackage{siunitx, hyperref}
\usepackage[keeplastbox]{flushend}

\setlist{nolistsep}

\def\BibTeX{{\rm B\kern-.05em{\sc i\kern-.025em b}\kern-.08em
    T\kern-.1667em\lower.7ex\hbox{E}\kern-.125emX}}

\begin{document}

\renewcommand{\algorithmicrequire}{\textbf{Input:}}
\makeatletter
\algrenewcommand\ALG@beginalgorithmic{\small}
\newcommand{\algmargin}{\the\ALG@thistlm}
\makeatother
\algrenewcommand{\algorithmiccomment}[1]{\footnotesize \ $// $#1  \small} 
\algnewcommand{\parState}[1]{\State%
	\parbox[t]{\dimexpr\linewidth-\algmargin}{\strut #1\strut}}

\algdef{SE}[DOWHILE]{Do}{doWhile}{\algorithmicdo}[1]{\algorithmicwhile\ #1}%

\newcommand*\circled[1]{\tikz[baseline=(char.base)]{
		\node[shape=circle,draw,inner sep=1pt] (char) {#1};}}

\title{Road Context-aware Intrusion Detection System for Autonomous Cars
\thanks{This work was done when Jingxuan Jiang was an intern in Singapore University of Technology and Design, Singapore.}\thanks{Chundong Wang is the corresponding author.}
}

\author{
	\IEEEauthorblockN{Jingxuan Jiang}
	\IEEEauthorblockA{\textit{Shandong University} \\
	\textit{jingxuan\_jiang@mail.sdu.edu.cn} \\
	}
	\and	
	\IEEEauthorblockN{Chundong Wang, Sudipta Chattopadhyay}
	\IEEEauthorblockA{\textit{Singapore University of Technology and Design} \\
		\textit{firstname\_lastname@sutd.edu.sg} \\
	}
	\and	
	\IEEEauthorblockN{Wei Zhang}
	\IEEEauthorblockA{\textit{Shandong University} \\
	 \textit{davidzhang@sdu.edu.cn}\\
	}
}

\maketitle

\begin{abstract}
Security is of primary importance to vehicles. 
The viability of performing 
remote intrusions onto 
the in-vehicle network has been manifested. 
In regard to unmanned 
autonomous cars, limited work has been done to detect intrusions 
for them while existing intrusion detection systems (IDSs) embrace
limitations against strong adversaries. 
In this paper, we consider the very nature of autonomous car and leverage the {\em road 
context} to build a novel IDS, named {\em R}oad context-{\em a}ware {\em IDS} (RAIDS).
When a computer-controlled 
car is driving through continuous roads, 
road contexts and genuine frames transmitted on the car's 
in-vehicle network should resemble a 
regular and intelligible pattern.
RAIDS hence employs a lightweight machine learning model to
extract road contexts from sensory information (e.g., camera images and
distance sensor values) that are used to 
generate control signals for
maneuvering the car. With such ongoing road context, RAIDS  
validates corresponding frames observed on the in-vehicle network.
Anomalous frames that substantially deviate from
road context will be discerned as intrusions. 
We have implemented a prototype of RAIDS with neural networks, and 
conducted experiments on a Raspberry Pi 
with extensive datasets and meaningful intrusion cases. 
Evaluations show that RAIDS significantly 
outperforms state-of-the-art IDS without using road context 
by up to 99.9\% 
accuracy
and short response time.

\end{abstract}


\input{intro2}

\input{background}

\input{motivation3}

\input{algo}
\input{evaluation}

\input{conclusion}

{
	\bibliographystyle{IEEEtran}	
	\bibliography{IDS}
}

\end{document}

%% file: intro2.tex
\section{Introduction}

Security is critical 
for vehicles.
A modern automobile embodies a protocol, like the 
 Control Area Network (CAN) bus~\cite{security:CAN:RESS-2011,car:CAN:RTSS-2017},
for in-vehicle communications among its 
electrical subsystems, including 
the steering wheel, brake, and engine,
each of which is monitored and controlled through an 
electronic control unit (ECU).
Researchers managed to manifest concrete 
intrusions to ECUs of manned vehicle 
to cause a breakdown or traffic 
accident~\cite{security:analysis:Security-2011,
security:fingerprint:Security-2016, IDS:DNN:Plos-2016, security:context-car:ESCar-2017}. 
Today,
many technology giants, startups, and academic researchers
are developing  
self-driving autonomous cars, which, undoubtedly,
demand particular care 
for security and safety~\cite{security:automobile:SP-2010, security:automated:TITS-2015}. 
The fatal accidents in Uber's and Tesla's autopilot tests have alerted people to 
such unmanned vehicles~\cite{car:Uber-accident:2018, car:Tesla-accident:2018}.
However, limited work has been done on designing
an intrusion 
detection system (IDS)
for the in-vehicle communications of autonomous car.
Existing
IDSs even have limitations against 
strong adversaries. Take the state-of-the-art
CIDS~\cite{security:fingerprint:Security-2016} for example.
In accordance with its knowledge of the
fingerprints (i.e., clock skews) 
of all ECUs, 
CIDS 
tracks down anomalies
when an original ECU stops sending frames 
or an ECU belonging to adversaries injects frames. Whereas,
CIDS should be oblivious 
of a compromised
ECU sending forged frames.
If a strong adversary can
manipulate an original ECU 
to deliver fake frames, CIDS would malfunction 
since the fingerprint of the ECU is not peculiar.
In other words,
such an attack model is beyond the capability of CIDS.

There is a fact that has not been considered in designing
IDS to protect in-vehicle network: 
all frames transmitted on the CAN bus are 
generated due to 
the decisions made by 
the vehicle driver and it is the {\em road context} that guides 
a driver to make those decisions. 
Human drivers have 
highly individualized experiences and
habits~\cite{car:human:HCI-2001}. 
They react differently to 
the same road context, like 
a stop sign or a road bend.
It is hence impractical to design an IDS with road context 
for manned vehicles. 
By contrast, an autonomous car is orthogonal to manned vehicles
concerning the very nature of `driver'. 
In a computer-controlled autonomous 
car, decisions are made by a well-trained self-driving
model upon dynamic road contexts obtained through multiple sensors~\cite{car:autonomous-behavior:ICCV-2015,car:Nvdia:arxiv-2016}.
Therefore, 
the road context and corresponding control signals, 
which eventually result in frames transmitted on the CAN bus, 
shall resemble a regular and intelligible
pattern for autonomous car. 
Given an intrusion with forged frames
upon continuous road contexts, 
a violation of the pattern must 
be perceivable.

Motivated by this observation, we develop
a holistic IDS for autonomous cars, 
namely \textbf{\em R}oad context-\textbf{\em a}ware
{\bf \textit{IDS}} ({\bf RAIDS}), 
to detect anomalous  
CAN frames issued by strong
adversaries. The main ideas of RAIDS, also the main contributions of this paper,
are summarized as follows.
\begin{itemize}[leftmargin=5mm,nosep]
	\setlength{\itemsep}{-\itemsep}
	\item RAIDS is a two-stage framework that mainly consists 
	of two neural networks 
	to extract road context from sensory information (e.g., images taken by cameras, 
	distances to front objects, 
	etc.) and validate the genuineness of CAN frames, respectively, for the purpose of 
	intrusion detection. Both neural networks are designed to be
	lightweight and efficient 
	regarding the computational resources of an in-vehicle embedded computing system.
	\item To extract road contexts, the neural network at the first stage of RAIDS processes camera images and other 
	sensory information that are concurrently used by the self-driving model to control the car.  
		The second stage of RAIDS is a binary classifier that 
	 verifies whether the frames observed on the CAN bus are abnormal or not
	 with regard to 
	 the extracted road contexts.
\end{itemize}
We have built a prototype of RAIDS.
A convolutional neural network (CNN) makes the backbone 
of RAIDS's first stage
for extracting and abstracting road contexts from camera images. 
The second stage of RAIDS 
mainly leverages linear layers 
to efficiently discern anomalous CAN frames with extracted road context. 
To evaluate RAIDS, we follow state-of-the-art work~\cite{security:IDS-LSTM:DSAA-2016} 
and implement an IDS that 
learns from historical CAN frames without considering road context.
We run both IDSs in a Raspberry Pi with extensive datasets. 
On defending two types of intrusions, i.e., abrupt and directed intrusions,
RAIDS substantially outperforms the IDS without road context by
up to 99.9\% 
accuracy and short response time.

The rest of this paper is organized as follows. 
In Section~2, we present the background and related works.
In Section~3, we show the motivation of this paper and attack model
we consider.
In Section~4, we detail the design and implementation of RAIDS. 
In Section~5, we present the evaluation results of testing RAIDS.
We conclude the paper in Section~6.

%% file: background.tex
\section{Background \& Related Works}~\label{sec:background}

\textbf{Intrusion Detection System}\hspace{2ex}
The necessity of intrusion detection system (IDS) is concrete for a vehicle.
A number of IDSs have been proposed targeting the in-vehicle
network~\cite{security:CAN:RESS-2011, security:structured:IAS-2010, security:entropy:IVS-2011, security:aspects:IVS-2011, security:fingerprint:Security-2016, security:IDS-time:ICOIN-2016}. 
An automobile is made of multiple electrical subsystems, each of which
has an electronic control units (ECU) to communicate with other subsystems
to control the vehicle. 
The Controller
Area Network (CAN) is the most widely employed protocol
for in-vehicle communications. ECUs encapsulate data
in CAN frames and put them on the CAN bus. 
A CAN frame contains no identity information of sender or receiver
for simplicity. The lack of identity in CAN frames facilitates
adversaries in fabricating hazardous messages. Worse,
modern vehicles are being connected to the outside world
via multiple channels, which leave exploitable attack vectors 
for adversaries to leverage.

One common mechanism of IDS is to observe and 
analyze frames transmitted among ECUs so that anomalous 
CAN frames can be detected. For example,
M\"uter and Asaj~\cite{security:entropy:IVS-2011}
proposed an entropy-based approach. They found that
frames on the CAN bus are more `regular' than ones
on the computer networks, which 
leads to a relatively low entropy for CAN frames.
As a result, injecting or 
dropping CAN frames should 
evidently increase the entropy of in-vehicle network 
and in turn expose an intrusion.
Song et al.~\cite{security:IDS-time:ICOIN-2016} 
worked in a similar fashion
but used the time interval between CAN frames to
inspect suspicious frames.
Taylor et al.~\cite{security:IDS-LSTM:DSAA-2016} 
emphasized on the data transmitted on CAN bus and
proposed a recurrent neural network (RNN) based anomaly detector.
Their RNN is trained with historical normal CAN bus data
so as to predict forthcoming frames and apprehend abnormal ones.
Whereas, their experiments for detecting anomalous frames were done
by manually flipping unused bits of data in a CAN frame 
to emulate an `unusual case'. 
Such a manipulation is irrational 
as skilled adversaries must have a good knowledge of transmitted
data and tend to fabricate meaningful but harmful frames. Wasicek
et al.~\cite{security:context-car:ESCar-2017} proposed to 
learn the `intra-vehicle context' by 
collecting the values of multiple sensors installed in 
a vehicle's subsystems
and building reference models to detect anomalies.
We note that their `context' is the internal context among a
vehicle's subsystems and has no relation to road context.
In the meantime, Cho and Shin~\cite{security:fingerprint:Security-2016}
proposed Clock-based IDS (CIDS) that
used the clock skew of ECUs to fingerprint them. Leveraging
the unique clow skew of each ECU, CIDS can not only detect
the occurrence of intrusions, but also locate the compromised ECU.

\textbf{Autonomous Car \& Neural Network}\hspace{2ex} 
There are different levels of automated driving~\cite{car:levels:2016}. 
In this paper, we consider an autonomous car that is computer-controlled 
most of the time except for emergency cases, such as an intrusion
to in-vehicle network.
Multiple sensors are installed to control an autonomous car,
including cameras, ultrasonic distance sensors, radar, etc. 
Such sensory information reflects and 
resembles real-world road context, and
advises the self-driving model to generate control signals.
Control signals are transformed to 
data encapsulated in frames transmitted on the CAN bus.

Numerical sensor values, 
like the distance to front objects,
are computer-readable and can be directly utilized by
the self-driving model of autonomous car.
The camera images, however, must be processed
to acquire high-level informative properties.
Nowadays, neural networks have emerged 
as the mainstream approach that deals with images
for self-driving. 
For example,
the convolutional neural network (CNN) has been proved to be effective 
in 
extracting image features to maneuver the autonomous car~\cite{DNN:CNN-road-segmentation:ECCV-2012,  DNN:CNN-features:CVPR-2014, car:resnet18:CVPR-2016, car:Nvdia:arxiv-2016}.
A CNN makes use of 
convolutional layers that 
apply multiple kernels
to extract embedded visual 
features.
A kernel is a small matrix of numbers. 
An image can be viewed as a large matrix
that comprises 
many small sub-matrices with the kernel size.
Convolutional layer {\em convolves} (i.e., slides) each kernel over 
sub-matrices of image to do matrix multiplication.
The output of a convolutional layer is thus a feature map
that bundles results of convolving multiple kernels.
In a CNN, feature maps of several convolutional layers, after
being computed through hidden  
layers for reduction of 
computations and avoidance of overfitting, 
will eventually make a vector that resembles  
the features per image. 
Such a feature vector is expressive and meaningful in image understanding~\cite{DNN:CNN-features:CVPR-2014,DNN:HDR-CNN:TOG-2017}. 

%% file: motivation3.tex
\section{Problem Formulation}~\label{sec:motivation}

\subsection{Motivation}
Most IDSs were developed with regard to human-driven vehicles.
Today, many technology giants, startups, and academic researchers
are developing
autonomous cars, which, undoubtedly,
demand particular care 
for security and safety, especially after 
two fatal accidents that successively happened 
in Uber's and Tesla's self-driving tests~\cite{car:Uber-accident:2018,car:Tesla-accident:2018}.

\begin{figure}[t]
	\centering
	\subfigure[Human-driven Car 1]{\includegraphics[width=0.305\columnwidth]{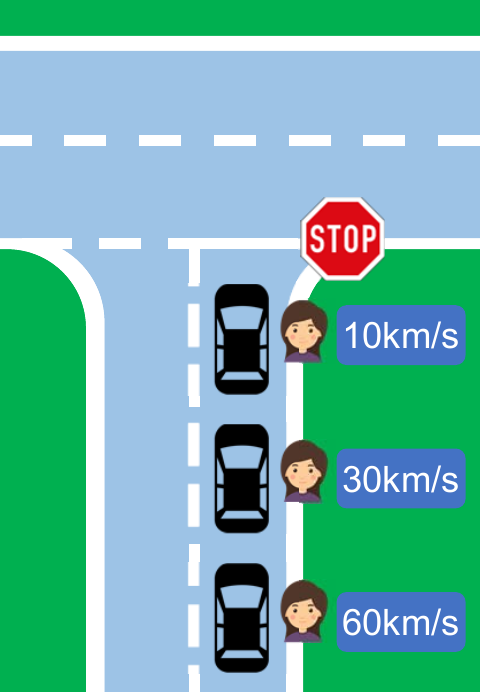}~\label{fig:stop1}}
	\subfigure[Human-driven Car 2]{\includegraphics[width=0.305\columnwidth]{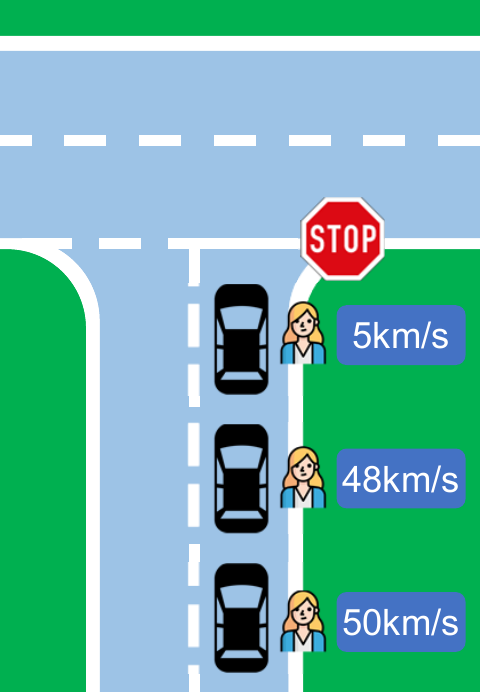}~\label{fig:stop2}}	
	\subfigure[Autonomous Car]{\includegraphics[width=0.305\columnwidth]{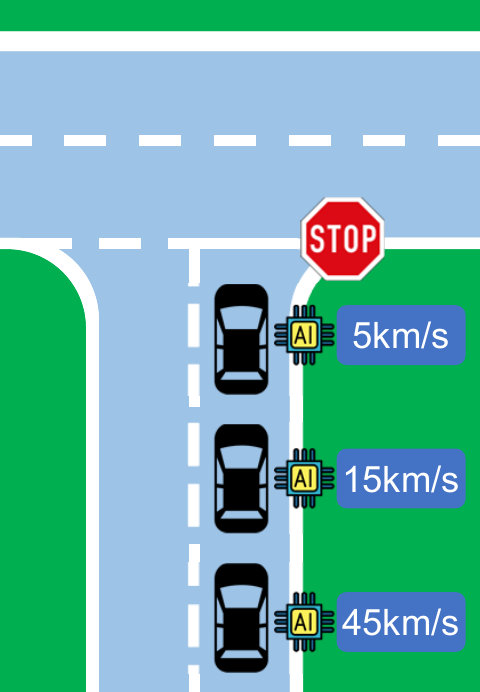}~\label{fig:stop3}}
	\vspace{-1ex}		
	\caption{An Illustration of Different Reactions of Human-driven and Autonomous Cars on a Stop Sign}\label{fig:stop}
	\vspace{-3ex}	
\end{figure}

Limited work has been done on detecting intrusions
to the in-vehicle network of autonomous car.
Worse,
state-of-the-art
IDSs embrace limitations against strong adversaries. 
For example,
CIDS~\cite{security:fingerprint:Security-2016} is able to detect intrusions
when a foreign ECU injects messages or 
an existing ECU stops sending messages based on its knowledge of
fingerprints (i.e., clock skews) 
of ECUs. However, if adversaries compromise
an ECU and use it to send fake messages, CIDS will be ineffective 
as the fingerprint is not
suspicious. In other words, CIDS should be oblivious of compromised
ECUs sending fake messages. Such intrusion cases
are beyond the capability of CIDS.

In practice,
frames transmitted on the CAN bus are generated when drivers 
encounter dynamic road contexts. 
Let us assume a stop sign as the road context ahead.
A driver must decelerate and eventually stop the car for a moment. 
The ECU of accelerator 
accordingly produces and delivers CAN frames with decreasing speed values. 
Figure~\ref{fig:stop1} and Figure~\ref{fig:stop2} illustrate
how two human drivers react when they move towards a stop sign. 
One driver gradually reduces speed. 
The other one decelerates only when 
being close to the stop sign. 
Because human drivers have different experiences and habits, 
they behave differently on the same road context and their reactions
entail diverse CAN frames~\cite{car:human:HCI-2001}.

When a car is controlled  
by a well-trained self-driving machine learning
model, however,
its behaviors should be smooth and stable. 
As shown by Figure~\ref{fig:stop3},
an autonomous car starts to reduce speed on spotting the stop sign and steadily
slows down in order to approach the stop line. 
This results in CAN frames with 
consistently decreasing speed values. 
Concretely, the road context (i.e., a stop sign) and
CAN frames (i.e., decreasing speed values) 
construct a regular and consistent pattern for autonomous car.
Assume that adversaries compromise the accelerator of autonomous car 
shown in Figure~\ref{fig:stop3} 
and continually put frames with non-decreasing 
speed values on the CAN bus. 
These abnormal frames 
are easy to be ruled out as they significantly deviate from the pattern
supposed for a stop sign.

\begin{figure*}[t]	
	\centering
	\scalebox{1.00}{\includegraphics[width=\textwidth]{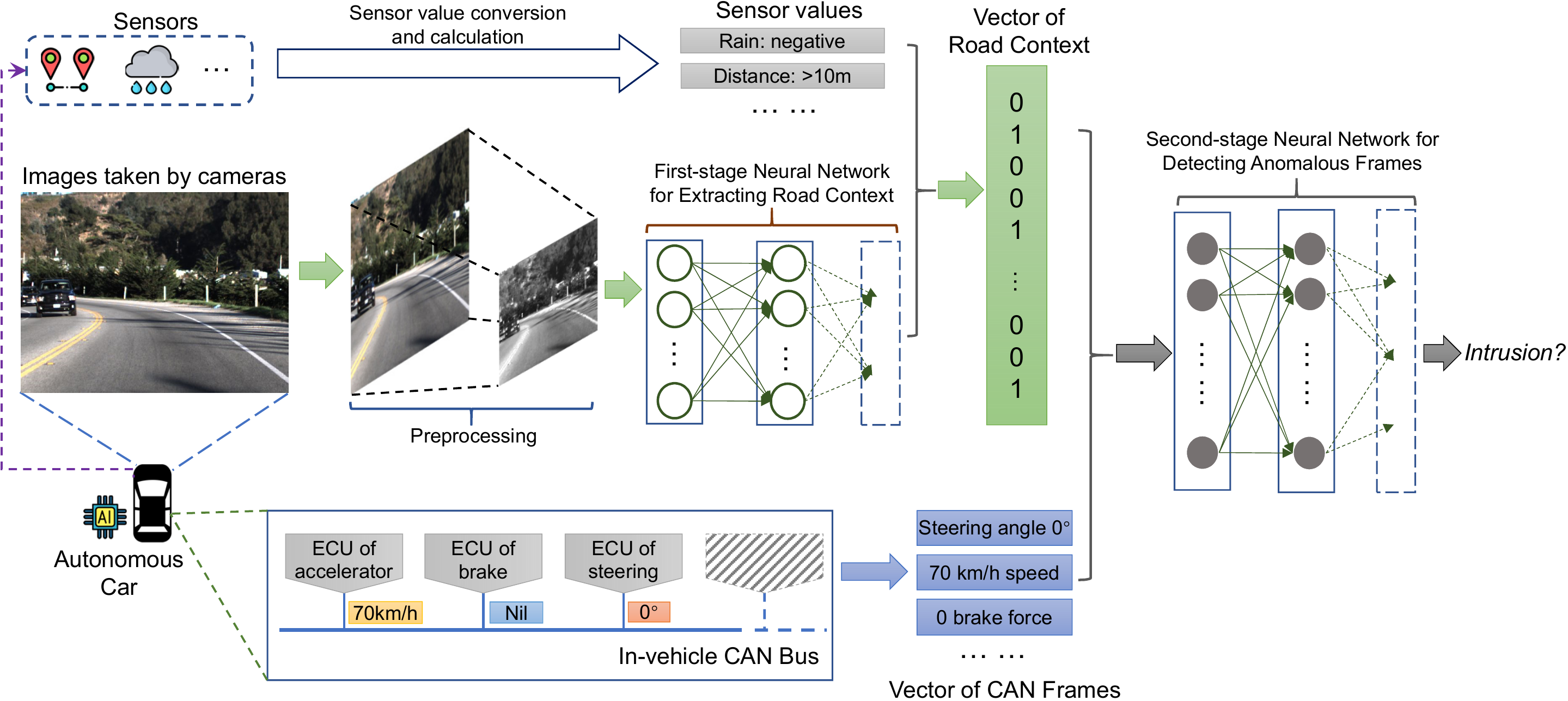}} 
	\vspace{-4ex}
	\caption{An Illustration of RAIDS's Architecture}~\label{fig:arch}
	\vspace{-3ex}	
\end{figure*}

To sum up, 
given a specific road context, 
the consequential CAN frames generated by an autonomous car 
are regular and predictable.
If we monitor ongoing road context and validate against
observed CAN frames, anomalous frames shall be detectable.
This observation motivates us to make a new 
\underline{R}oad context-\underline{a}ware \underline{IDS} (RAIDS) for autonomous car.

\subsection{Attack Model}\label{sec:attack}
A number of network connections exist in an autonomous vehicle 
to help it communicate with the outside world. These connections yet
provide attack vectors for adversaries to exploit. 
We assume that strong adversaries further have good knowledge of in-vehicle network, 
including the format and frequency of CAN frames issued by an ECU, and
also manage to force an ECU to encapsulate and send their data
in CAN frames.
With such knowledge, adversaries are able to remotely access and manipulate
critical ECUs of an autonomous car, such as the steering wheel, brake, and accelerator.
In this paper, we consider an attack model that is beyond 
the capability of state-of-the-art IDSs, i.e., {\em forgery attack}.

The process of a forgery attack is as follows.
Once adversaries compromise an ECU on the in-vehicle network,
they first intercept the normal frames sent and received by the ECU in 
order to study the ECU's behavior and data format. Then
adversaries start forging and sending CAN frames strictly with the original frequency. 
Nevertheless,  
the data put in forged frames is made either inappropriate or opposite 
due to the malicious intentions of adversaries.
For example, upon a left turn, adversaries may replace CAN 
frames of the steering wheel with
right turn angles 
so as to wreck the car.

Forgery attack has two variants.
\begin{itemize} [leftmargin=5mm,nosep]
	\setlength{\itemsep}{-\itemsep}
	\item \textit{Abrupt intrusion}: adversaries abruptly place anomalous 
	CAN frames with abnormal data at a random time to cause a disorder. 
	\item \textit{Directed intrusion}: adversaries monitor the road context at runtime 
	and, upon a specific scenario, like a road bend or traffic light, place anomalous
	CAN frames that significantly violate the road context. 
\end{itemize}
Both types of intrusions are harmful. 
Assuming that
the self-driving model is unaware of any intrusion,
the consequence of directed intrusion is more detrimental as
the CAN frames imposed by it are to inflict a sudden flip 
to the vehicle's state, like the aforementioned right turn upon a left turn.

%% file: algo.tex
\section{RAIDS}

\subsection{Overview of RAIDS}

As its name suggests,
the essence of RAIDS is to leverage the ongoing road context to 
validate whether CAN frames on the in-vehicle network
are normal or not for an autonomous car. 
If the CAN frames closely match the corresponding road context, 
RAIDS deems that there is no security threat.
Otherwise, RAIDS will report the occurrence of intrusion.

\begin{figure}[t]
	\centering
	\subfigure[The Road Context of a Straight Road]{\includegraphics[width=\columnwidth]{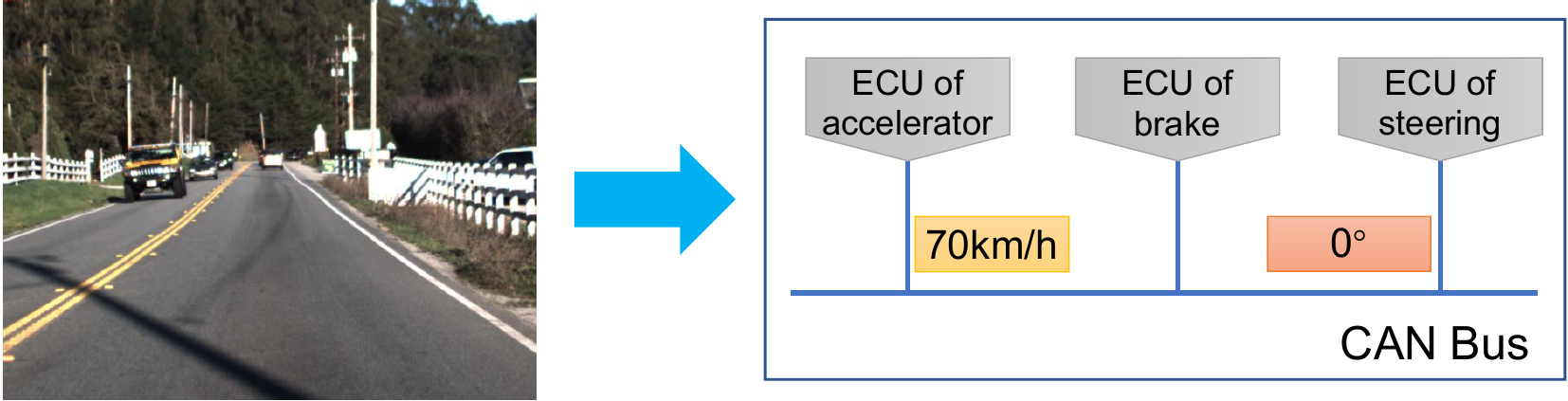}~\label{fig:straight}}
	\subfigure[The Road Context of a Road Bend]{\includegraphics[width=\columnwidth]{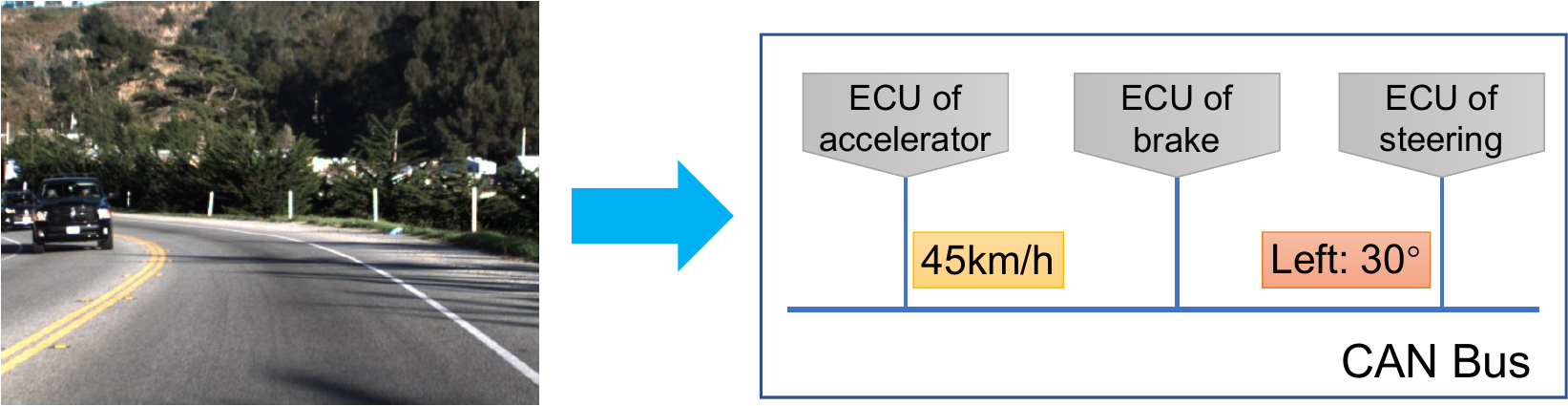}~\label{fig:bend}}	
	\vspace{-2ex}			
	\caption{An Illustration of the Impact of Road Context on CAN Frames}
	\label{fig:context}
	\vspace{-3ex}			
\end{figure}

\autoref{fig:arch} illustrates the architecture 
of RAIDS\footnote{The images used in \autoref{fig:arch},~\autoref{fig:context}, ~\autoref{fig:kernels},
and \autoref{fig:cnn} are from the dataset of Udacity self-driving challenge~\cite{car:udacity-dataset} under MIT license.}. 
As shown by the leftmost of~\autoref{fig:arch}, 
the ongoing road context is reflected by a variety of sensory information, such as the
distances to surrendering objects, 
the water level implying a rainy slippery road, 
and camera images showing the front scene. Numerical
sensory information is computer-readable 
while camera images must be processed.
The self-driving model depends on sensory information to decide
how to maneuver the autonomous car.
In the meantime, sensory information is also  
delivered to RAIDS. 
RAIDS is mainly composed of two neural networks.
One neural network is responsible for processing the camera images
which cannot be instantly utilized.
The image
is first preprocessed through techniques like 
normalization and centering. 
Then RAIDS uses one neural network, as shown at the central part of~\autoref{fig:arch},
to extract and abstract image features.
These image features will be concatenated with other numerical
sensory information to make a vector of road context.
On the other hand, as illustrated by the lower half of~\autoref{fig:arch},
the self-driving model produces control signals upon the  
sensory information, which
eventually conveys a number of CAN frames transmitted on
the in-vehicle network. 
These CAN frames are formulated into another vector that
is fed along with the vector of road context as two inputs 
to the second neural network of RAIDS. As shown by the rightmost part of 
~\autoref{fig:arch}, 
with well-trained parameters learned from 
historical road contexts and CAN frames,
the second neural network shall tell whether 
abnormal frames emerge on the CAN bus or not.
RAIDS immediately informs the self-driving model 
once any anomaly is detected.

If an intrusion is reported,
RAIDS suggests that
the self-driving model should 1) first disable external
network connections to block remote adversaries, 
2) stop the vehicle for emergency if possible,
and 3) raise a switch request to human driving.
These steps aim to mitigate the impact of intrusions.

\subsection{Road Context}\label{sec:context}
We define the road context as 
{\em the information an autonomous car is encountering 
when it is cruising}. In summary, the road context includes 
but not limited to, 
1) the road conditions, like lane line, cross, bend, joint, and fork of roads,
2) the traffic light, pedestrians, vehicles, obstacles, bumps, and pits 
around the autonomous car, 
3) the weather conditions, such as rain, fog, cloud, and snow, 
and 4) the sunrise, sunset, night and tunnel lights.
These road contexts are perceived by multiple sensors installed in
the autonomous car, such as cameras, ultrasonic distance sensors, water sensor, etc.

The road context determines control signals made by the vehicle driver,
which, with regard to the computer-controlled autonomous car, is the self-driving 
model. Different road contexts shall entail different signals, 
which in turn generate different frames on the in-vehicle CAN bus.
~\autoref{fig:context} instantiates 
the impact of road curves
on the control signal. 
As shown in Figure~\ref{fig:straight},
on a highway that is continuously straight, 
the self-driving model demands the autonomous car to move straightforward 
and run at a velocity of 70km/h.
By contrast, upon a road bend as illustrated in Figure~\ref{fig:bend}, 
the framework shall decrease the car's velocity to 
45km/h and turn to the left with an angle of \ang{30}.
Assuming that on the road shown in 
Figure~\ref{fig:bend}, 
a frame with a steering angle of \ang{0} for moving straightforward,  
rather than the rational frame of left turn with \ang{30}, emerges
on the CAN bus, an intrusion should have 
taken place because the anomalous frame is not congruent 
with the ongoing road context.

CAN frames are ever-changing due to dynamic road contexts from time to time.
The tight relation between road context and CAN frames indicates that
road context must be exploited as a crucial parameter to detect intruions initiated onto
the in-vehicle network.
We hypothesize that the self-driving
model of autonomous car is intact and always
makes wise and regular decisions upon 
dynamic road contexts. In other words,
an autonomous car is a contrast to 
manned vehicles in which human
drivers may behave inconsistently 
from time to time even regarding the same road context~\cite{ car:human:HCI-2001}.
In addition,
we note that the focus of this paper is on detecting security threats imposed 
by adversaries onto ECUs and in-vehicle network. Readers may refer to other
studies for the vulnerability exploration of deep learning models
that drive a vehicle~\cite{car:DeepXplore:SOSP-2017, car:DeepTest:ICSE-2018, DNN:backdoor:SP-2019, DNN:pixel-attack:TEC-2019}. 

\input{extract}

\subsection{Intrusion Detection with Road Context}\label{sec:detect}

After obtaining the vector of road context 
and CAN frames corresponding to the road context, 
we can establish a model between them by learning over historical
records of road contexts and CAN frames. 
With the model, RAIDS validates whether observed CAN frames approximately 
match the newly-arrived image and sensor values.
By doing so, RAIDS determines 
the genuineness of CAN frames. 
A substantial discrepancy would 
lead to a report of intrusion.

\begin{table*}[tb]
	\centering
	\caption{The Datasets and Intrusions Used to Evaluate RAIDS}\label{tab:dataset}
	\vspace{-1ex}
	\resizebox{\linewidth}{!}{
		\begin{tabular} {l|l|l|l|l}
			\hline
			\multirow{3}{*}{Datasets} & \multirow{3}{*}{Sources} & Genuine Steering & \multirow{3}{*}{Manipulations of Abrupt Intrusion} & \multirow{3}{*}{Manipulations of Directed Intrusion} \\ 
			&& Angle Ranges in &&\\
			&& Radian &&\\		
			\hline\hline
			Udacity & Udacity self-driving challenge~\cite{car:udacity-dataset} & [-2.05, 1.90] & Randomly select 30\% images and for an image, add or sub- &  \\ \cline{1-3}
			Udacity\_sim & Udacity simulator~\cite{car:udacity-sim-dataset} & [-0.94, 1.00] &tract a random value in [0.1, 0.9] to its corresponding angle. & Select the largest 15\% and smallest  \\ \cline{1-4}
			\multirow{2}{*}{Apollo} & \multirow{1}{*}{Road Hackers platform in Baidu} & \multirow{2}{*}{[-0.38, 0.21]} &Randomly select 30\% images and for an image, add or sub- & 15\% angles. Flip the sign of a selected \\  
			& Apollo Project~\cite{car:apollo}&&tract a random value in [0.08, 0.5]to its corresponding angle. & angle if its absolute value is larger  \\ \cline{1-4}		
			Chen\_2017 & Recorded by Sully Chen in 2017~\cite{car:Chen}& [-1.99, 0.55] & Randomly select 30\% images and for an image, add or sub-& than 0.3; otherwise, add or subtract a  \\ \cline{1-3}
			Chen\_2018 & Recorded by Sully Chen in 2018~\cite{car:Chen} & [-2.01, 0.68]& tract random value in [0.2, 0.9] to its corresponding angle.& random value in [0.5, 1].\\ \cline{1-4}
			\multirow{2}{*}{Comma.ai} & \multirow{2}{*}{Comma.ai highway driving~\cite{car:commaai}} & \multirow{2}{*}{[-1.64, 1.29]} & Randomly select 30\% images and for an image, add or sub- & \\ 
			&&&tract a random value in [0.25, 1] to its corresponding angle. & \\ \hline
		\end{tabular}
	}
\end{table*}
\normalsize

As historical records of sensory information and CAN frames are known
as normals, how to detect intrusions on the in-vehicle network
regarding road context turns to be 
a problem of developing a supervised learning model 
to check CAN frames upon 
forthcoming sensory information. Assume that
$N$ items of sensory information
are used for training. The $i$th sensory information ($0\leq i < N$) 
has a vector of road context $r_i$ obtained through the first stage of RAIDS.
Still for the $i$th sensory information, CAN frames issued by ECUs like
steering wheel, accelerator, and brake 
have been collected and presented in a vector $c_i$.
In the perspective of supervised learning, we can
make a label $\lambda_i$ (`1' for normal and `0' for anomaly) for 
$r_i$ and $c_i$.
Given these $N$ tuples forming a training dataset, i.e., 
\begin{equation}~\label{eq:set}
\begin{matrix}
\{\langle r_0, c_0, \lambda_0 \rangle, 
\langle r_1, c_1, \lambda_1 \rangle, 
..., \langle r_i, c_i, \lambda_i \rangle, ...,\\
 \langle r_{N-1}, c_{N-1}, \lambda_{N-1}\rangle\},
\end{matrix}
\end{equation}	
RAIDS's supervised learning
attempts to seek out a model,
\begin{equation}
	g(r_i, c_i) \approx \lambda_i, (0 \leq i < N).
\end{equation}
Or put in another way,
\begin{equation}
g: \bm R \times \bm C \rightarrow \bm \Lambda, 
\end{equation}
in which $\bm R$, $\bm C$, and $\bm \Lambda$
 are the domain spaces of road contexts, CAN frames, and labels, respectively.

The function $g$ should be one that
the best describes the relationship between $\bm R$, $\bm C$, and $\bm \Lambda$. 
For a new element of $\bm R \times \bm C$, which will be forthcoming 
road context $r_x$ and its corresponding vector of CAN frames $v_x$, 
RAIDS computes $g(r_x, v_x)$ and obtains a label $\lambda_x$. 
If $\lambda_x$ is `0', RAIDS deems there would be no intrusion at that moment.
Otherwise, RAIDS informs the self-driving model of a
possible intrusion on the CAN bus.
Consequently, the second stage of RAIDS is  
formatted as a problem of 
binary classification.
In the prototype of RAIDS, we build a classifier that is mainly composed of
two linear layers. 
There are two reasons to do so.
First, $r_i$, $v_i$, and $\lambda_i$
are numerical vectors. Linear layers are sufficient to
speculate their relationship. 
Second, linear layers bring about relatively simpler
computations, which are especially efficient concerning the
response time of IDS and the computational resources
of an embedded computing system.

\subsection{Training and Testing}

As mentioned, the backbones of the 
first and second stages are a CNN and a binary classifier, 
respectively. 
They are based on Keras~\cite{python:keras} and PyTorch~\cite{python:pytorch} frameworks.
We follow an end-to-end learning fashion~\cite{car:Nvdia:arxiv-2016, car:DeepLanes:CVPR-2016} to train RAIDS.
The loss function is BCELoss (Binary Cross-Entropy loss) provided by PyTorch 
 for binary classification~\cite{python:pytorch-bceloss}.
We would use six datasets to evaluate RAIDS (more details can be found in Section~\ref{sec:evaluation}).
In each dataset, we use 70\% images and CAN frames for training while
the 30\% remainders are used for the purpose of testing. 
We note that because datasets have different image sizes,
RAIDS would have different implementation variants to deal with respective datasets.

%% file: extract.tex
\subsection{Extracting Road Context}\label{sec:extract}

We first need to extract the road context in order to  
leverage it for intrusion detection.
As a matter of fact, 
most of the road context
is reflected by the front scene that autonomous car is facing,
including the aforementioned road condition, traffic light and weather.
Such a scene is tracked by multiple sensors.
Numerical sensory information, like the distance to front objects, 
can be directly utilized by RAIDS 
since they are both human- and computer-readable.
The images captured by cameras, 
however, need to be converted into a
format that RAIDS can deploy.
As a result,
the difficulty of obtaining road context lies in  
how to process camera images.

\begin{figure}[t]	
	\centering
	\scalebox{1.00}{\includegraphics[width=\columnwidth]{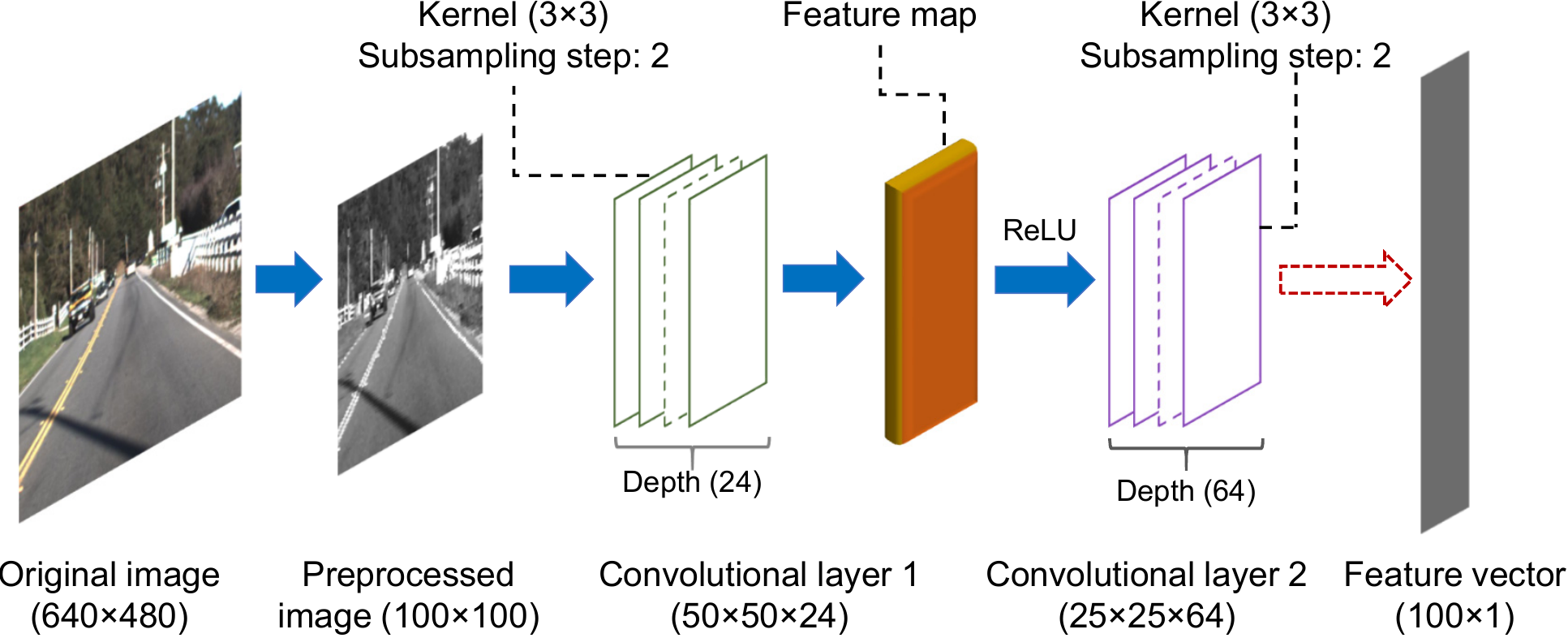}} 
	\vspace{-3ex}
	\caption{The Architecture of RAIDS's CNN}~\label{fig:kernels}
	\vspace{-3ex}			
\end{figure}

As mentioned in Section~\ref{sec:background}, 
the feature vector obtained in a deep neural network is a promising abstraction
of road context contained in an image
for RAIDS to leverage for intrusion detection.
Concretely, we construct a CNN 
as the backbone of the first stage of RAIDS 
to extract and abstract the road context from camera images. 
~\autoref{fig:kernels} sketches the architecture of the CNN
employed by RAIDS to process images. 
It mainly consists of two convolutional layers.
After being preprocessed, the first convolutional 
layer would apply 24 kernels, each of which is 3$\times$3 with a subsampling
step\footnote{A subsampling of 2 means that the convolutional layer 
	moves each kernel by 2, rather than 1, 
	when sliding over the large matrix of image,
	to reduce the dimensionality
	of feature map but without losing
	important information of the image. So the size
	of feature map for one kernel is $\frac{100}{2}\times\frac{100}{2} = 50\times50$.} of 2,
to generate a feature map (50$\times$50$\times$24). 
This feature map goes through a rectified linear unit (ReLU),
which is the activation function used 
in our implementation,
and then reaches
the second convolutional layer.
The second convolutional layer applies 64 kernels, each of which also has
3$\times$3 size with a subsampling step of 2. 
The second feature map is hence 25$\times$25$\times$64, and
would entail a feature vector of 100$\times$1 after passing  
one dropout layer and two dense layers.

\begin{figure}[t]
	\centering
	\subfigure[Straight Road]{\includegraphics[width=\columnwidth]{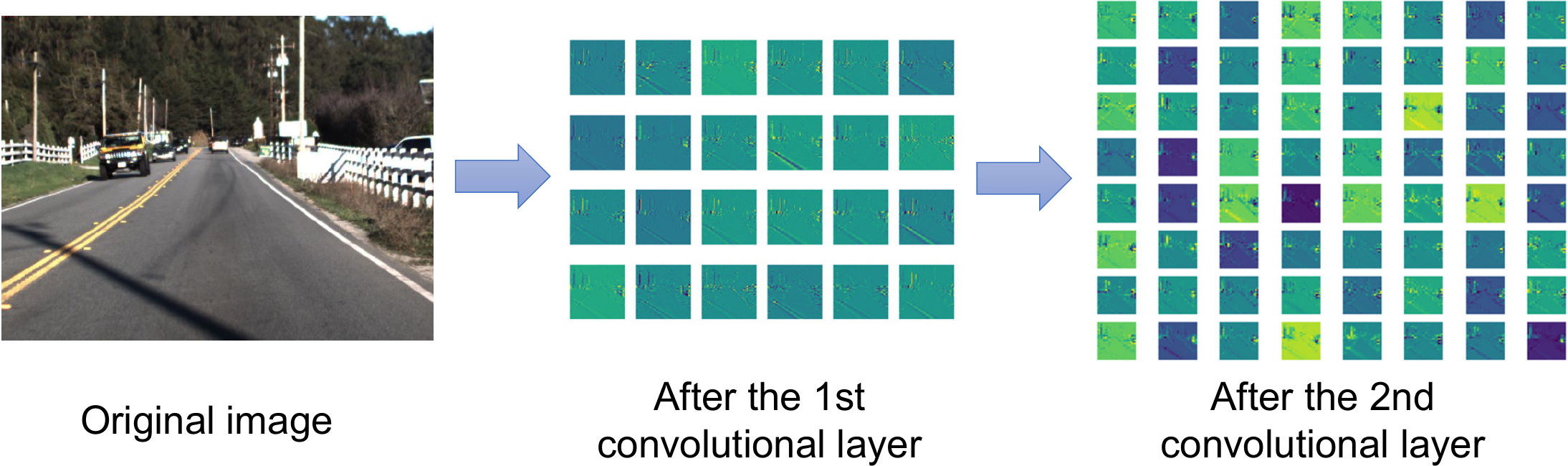}~\label{fig:cnn-straight}}
	\subfigure[Road Bend]{\includegraphics[width=\columnwidth]{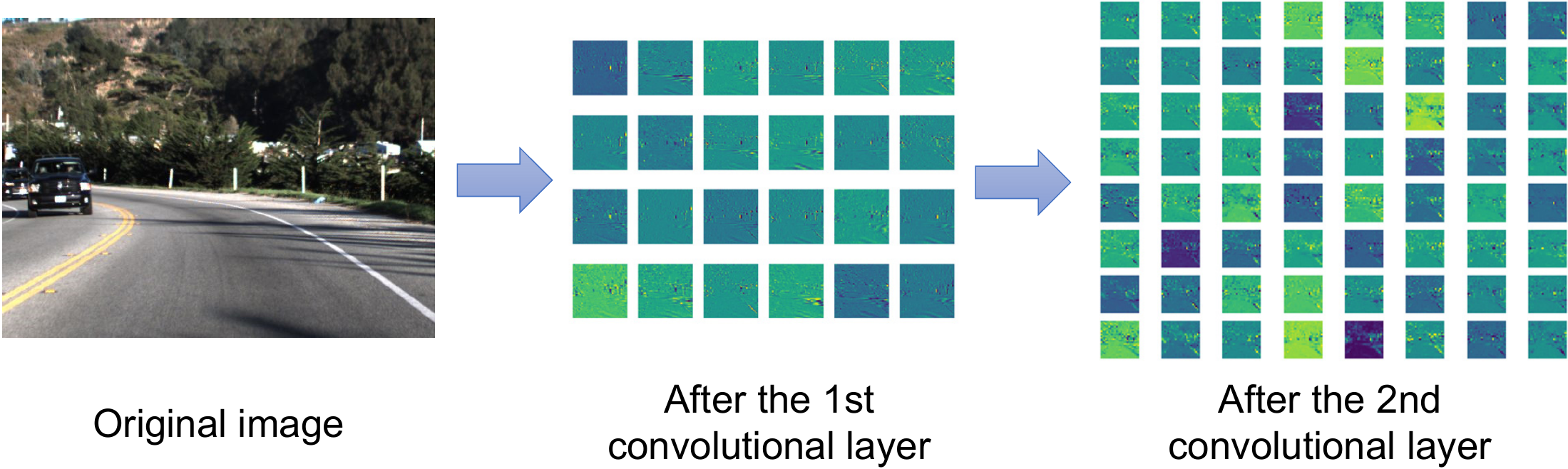}~\label{fig:cnn-left}}	
	\vspace{-2ex}			
	\caption{An Illustration of Extracting Features from Images by RAIDS's CNN}
	\label{fig:cnn}
		\vspace{-3ex}			
\end{figure}

\autoref{fig:cnn} exemplifies the feature maps visualized
after two convolutional layers for two images from Udacity dataset~\cite{car:udacity-dataset} when they are
being processed by the CNN of RAIDS.
A comparison between Figure~\ref{fig:cnn-straight} and Figure~\ref{fig:cnn-left}
confirms that different road contexts lead to different intermediate features.
In the end, the feature vector of each image would be 
assembled with numerical sensor values into 
a new vector as one of the inputs to the second stage of RAIDS.

The CNN extracting image features for RAIDS works synchronously
with the self-driving model of autonomous car, because 
RAIDS should 
verify CAN frames produced by the self-driving 
model on the same road context.
RAIDS's CNN is simpler than the self-driving model's. 
The reason is twofold. First, the self-driving model does not terminate with image features
but has to accordingly do further computations to determine control signals
subsuming left or right steering with a degree, acceleration with a velocity, 
brake with a force, etc.
Second, the feature vector generated by the CNN of RAIDS can be coarse-grained as long as
 they are sufficiently accurate for detecting intrusions at the second stage of RAIDS.

%% file: evaluation.tex
\section{Evaluation}~\label{sec:evaluation}

We have performance evaluations to answer three questions.
\begin{enumerate}[nosep,label*={\bf Q\arabic*}.]
	\setlength{\itemsep}{-\itemsep}
	\item Does RAIDS achieve high accuracy in intrusion detection? Is the performance of RAIDS stable over different datasets?
	\item How is the efficacy of RAIDS? Does it cost reasonable response time to detect an intrusion in an embedded computing system?
	\item Is RAIDS effective in detecting intrusions under more difficult road contexts?
\end{enumerate}

\subsection{Evaluation Setup}

\textbf{Datasets}\hspace{2ex} We have used six datasets from 
five sources. Their descriptions are presented in~\autoref{tab:dataset}.
Except Udacity\_sim that includes images recorded in a synthesized 
simulator, all other datasets were collected in the real world. 
These datasets contain a large number of records with images and 
corresponding data conveyed in CAN frames.

\textbf{Road Context}\hspace{2ex} As mentioned in Section~\ref{sec:context},
the road context is a broad concept and covers many aspects of the environment where
a vehicle is cruising. Concretely, 
we would place emphasis on the road conditions reflected by camera images, such as
lane lines, road bends, and turns. 
There are two reasons to do so.
First, not all datasets provide extra numerical sensor values.
Second, less sensory information imposes more challenges in
precisely obtaining road context. 

\begin{figure}[t]		
	\centering
	\subfigure[The Curve of Steering Angle]{\includegraphics[width=\columnwidth]{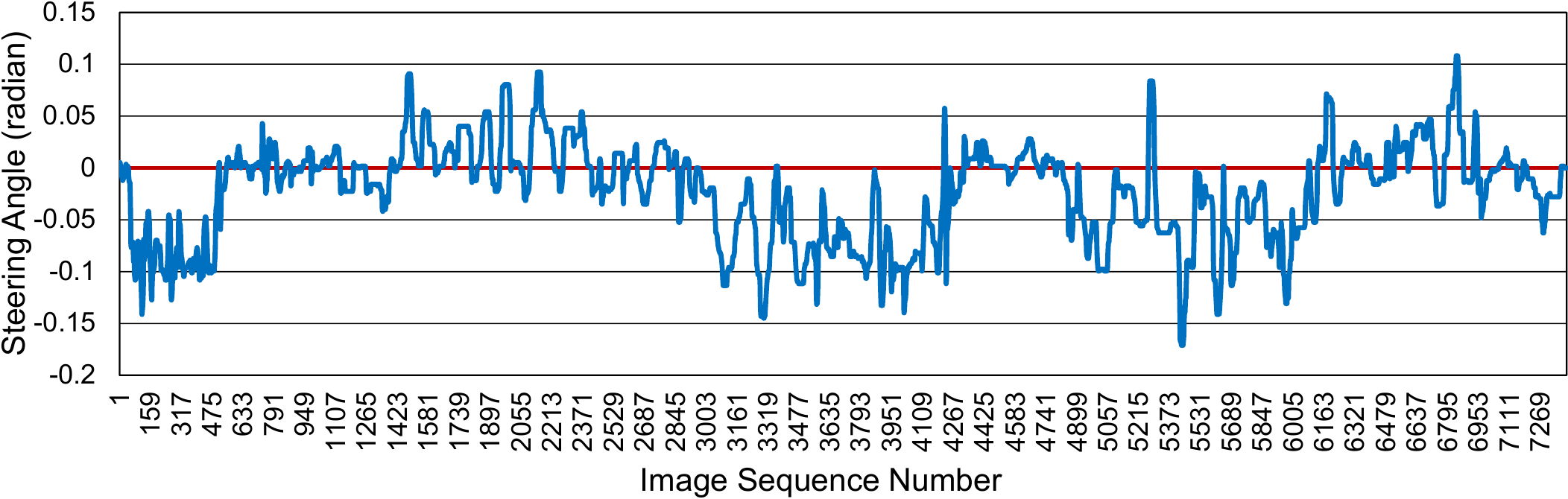}~\label{fig:angle}}
	\vfill
	\subfigure[The Curve of Vehicle Speed]{\includegraphics[width=\columnwidth]{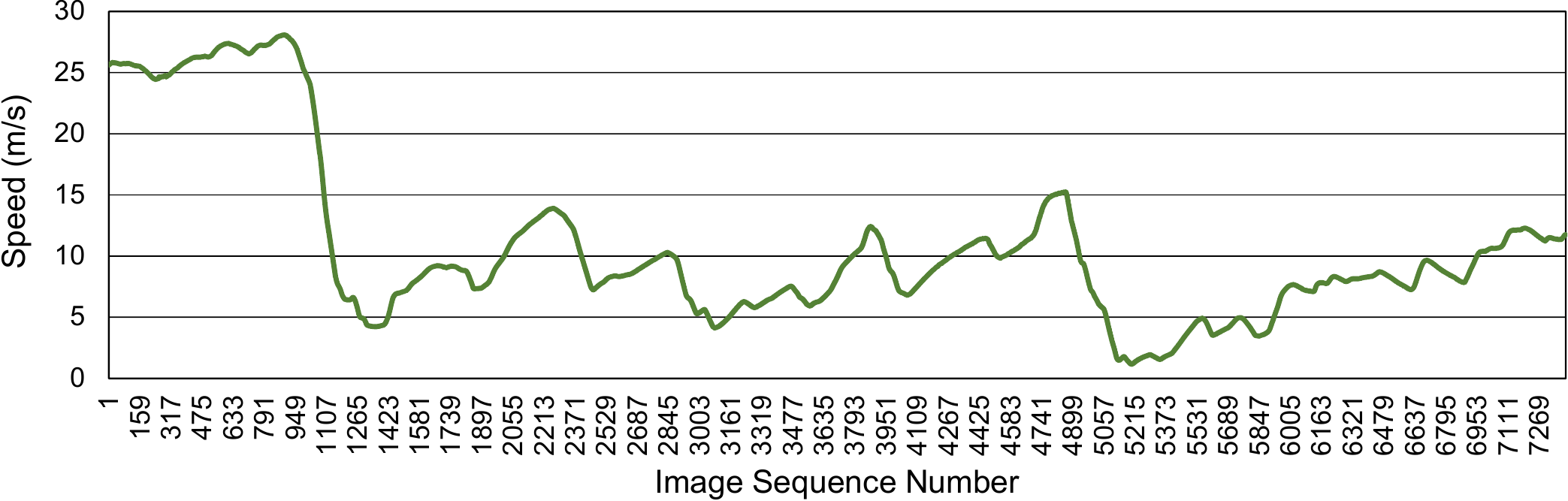}~\label{fig:speed}}
	\vspace{-2ex}		
	\caption{The Curves of Steering Angle and Vehicle Speed over Time with Udacity's HMB\_6 Dataset}\label{fig:angle-speed}
	\vspace{-2ex}	
\end{figure}

On the other hand, at the standpoint of adversaries, we would 
focus on intrusions onto the steering wheel. The reason is threefold.
First, the steering angle is one vital control signal for
autonomous car and attracts wide attention for research.
Second, the steering angle is  
ever-changing along the road 
while the control signals from accelerator and brake remain 
relatively stable for a moving vehicle. 
~\autoref{fig:angle-speed} sketches two curves for the steering angle
and vehicle speed at runtime, respectively, with one Udacity sub-dataset (HMB\_6).
It is evident that the curve of vehicle speed is much smoother 
than that of steering angle. Therefore, 
an intrusion to compromise steering angle 
is more difficult to be detected. 
Third, some datasets, like Chen\_2017 and Chen\_2018,
only include the runtime values of steering angle.

\begin{figure*}[t]
	\centering
	\subfigure[Abrupt Intrusion]{\includegraphics[width=\columnwidth]{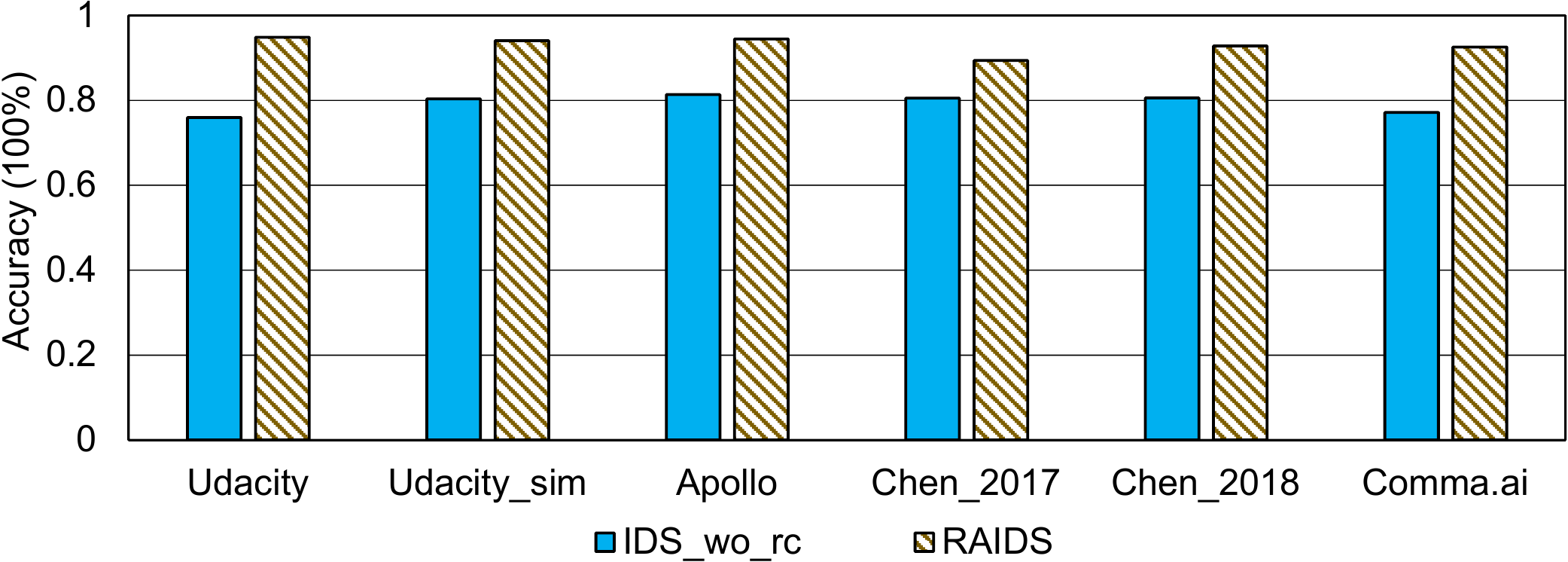}~\label{fig:random-accuracy}}
	\subfigure[Directed Intrusion]{\includegraphics[width=\columnwidth]{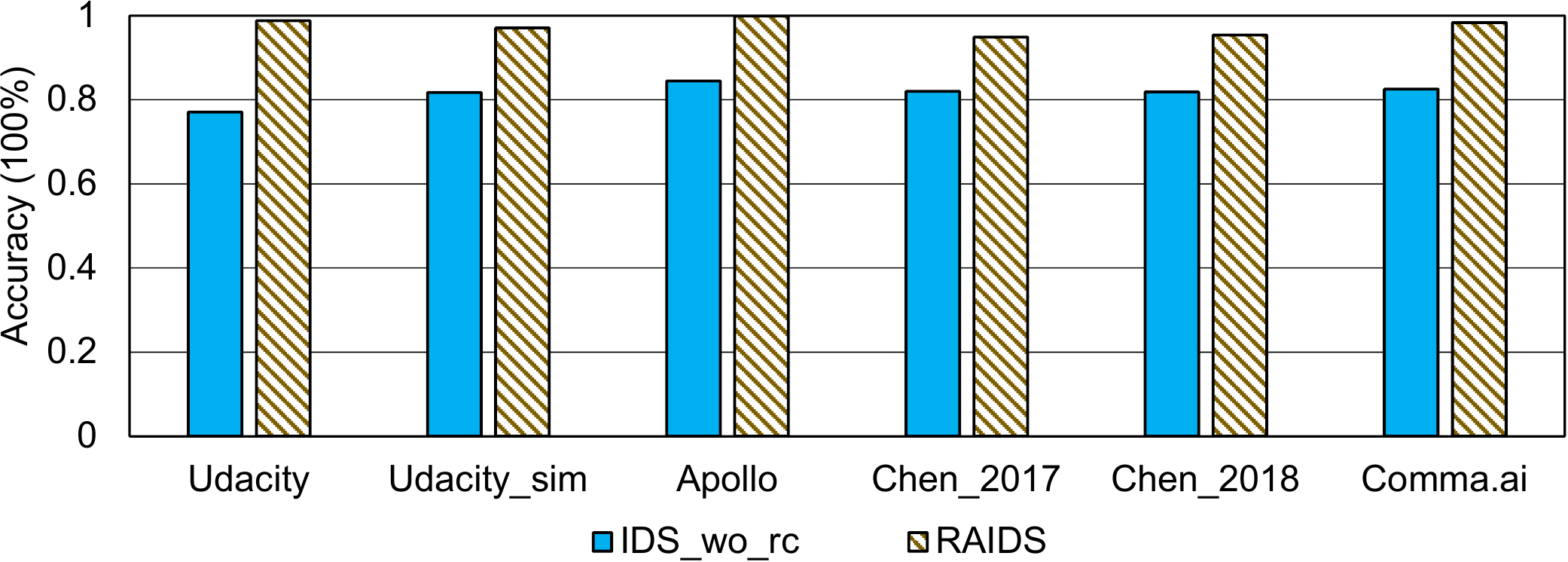}~\label{fig:directed-accuracy}}		
	\vspace{-2ex}		
	\caption{The Accuracies of Two IDSs under Two Types of Intrusions with Six Datasets}\label{fig:accuracy-summary}
	\vspace{-2ex}	
\end{figure*}

\textbf{Intrusions} \hspace{2ex} We consider forgery attack (cf.~Section~\ref{sec:attack}) in evaluation.
We have performed abrupt and directed intrusions 
to the values of steering angle with each dataset. 
In contrast to existing works that encapsulated meaningless data in CAN frames or
fabricated artificial CAN frames, our intrusions would generate a CAN frame
that should be with an allowable value but does not match the ongoing 
road context. In other words, such forged CAN frames emerge 
alongside inappropriate road contexts. The rightmost two columns of
~\autoref{tab:dataset} brief how we manipulate steering angles
to produce abrupt and directed intrusion cases.
Note that the datasets have different ranges for
steering angles. For example, for the HMB\_6 sub-dataset of Udacity
shown by Figure~\ref{fig:angle},
steering angles fall in [-0.17, 0.11] while the range for Chen\_2018
is [-1.99, 0.55]. For each dataset, we hence apply appropriate values
to modify the steering angles so as to make intrusion 
cases which are not trivial to be perceived.

\textbf{Competitor}\hspace{2ex}
Some existing IDSs like CIDS are ineffective against forgery attacks.
We implement an IDS without considering 
the road context (referred to as IDS\_wo\_rc)
for comparison. It 
is identical to the start-of-the-art IDS proposed by Taylor et al.~\cite{security:IDS-LSTM:DSAA-2016}. In brief,
IDS\_wo\_rc depends on learning CAN frames with RNN
to determine whether an arriving frame contains
genuine data or not. For IDS\_wo\_rc, 70\% of each dataset is used 
for training while 30\% is for testing, the same as what we do with RAIDS.
For both RAIDS and IDS\_wo\_rc,
training is performed in a Linux server while
testing is done in a Raspberry Pi 3 Model B+~\cite{bench:rpi3:2018}.
Python 3.5 is installed in 
both the server and Raspberry Pi.
We assume that an embedded system being with Raspberry Pi's computing powers  
exists within the in-vehicle network gateway~\cite{security:gateway:SAFECOMP-2008}
where the IDS resides for autonomous car.

\textbf{Metrics}\hspace{2ex} The main metric we use to compare RAIDS and IDS\_wo\_rc
is the detection accuracy in testing, i.e., the ratio of 
detected normal and intrusion cases against overall cases. 
A higher detection accuracy means a better effectiveness of IDS.
We also measure the ratios of undetected intrusion cases as well as false alarms 
 by which an IDS wrongly labels a normal CAN frame to be anomalous.  
To study the efficiency of RAIDS, we record 
the average and maximum response time RAIDS spends in processing all
cases of a dataset.

\begin{figure*}[t]
	\centering
	\subfigure[Udacity under Abrupt Intrusion]{\includegraphics[width=0.485\columnwidth]{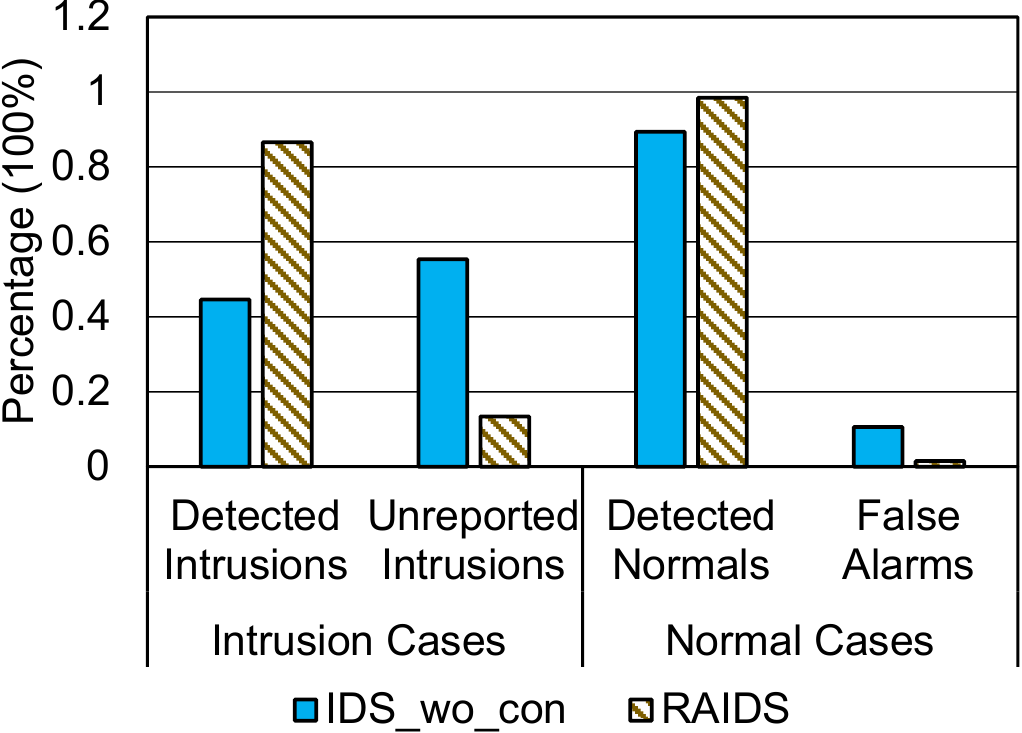}~\label{fig:false-random-udacity}}
	\subfigure[Udacity under Directed Intrusion]{\includegraphics[width=0.485\columnwidth]{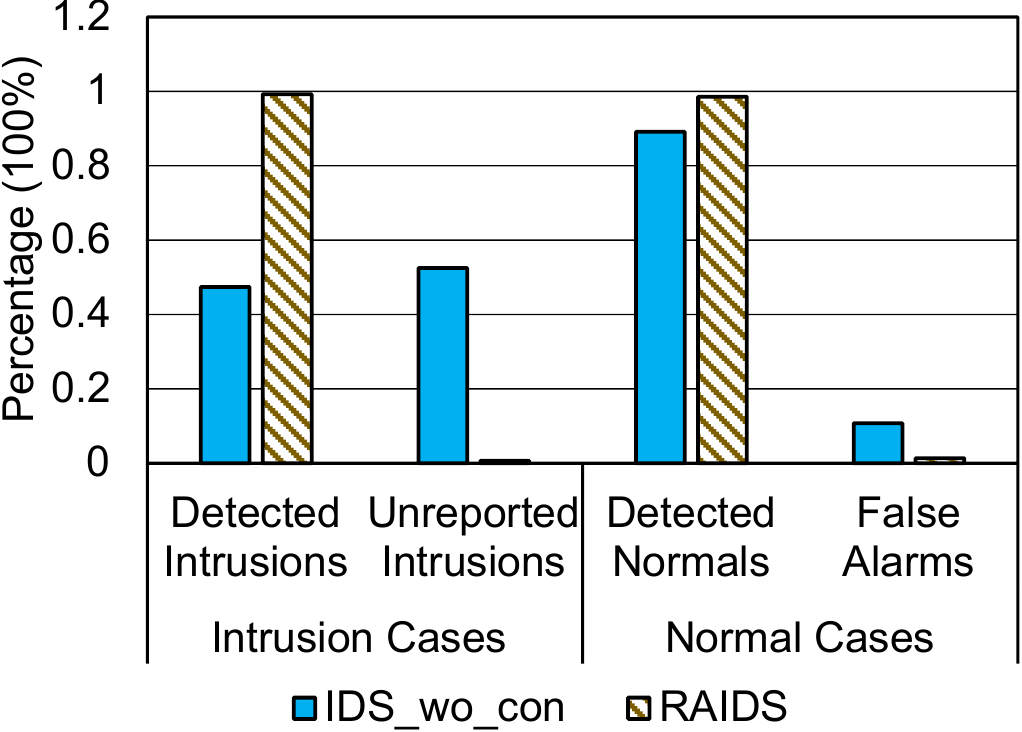}~\label{fig:false-directed-udacity}}
	\subfigure[Udacity\_sim under Abrupt Intrusion ]{\includegraphics[width=0.485\columnwidth]{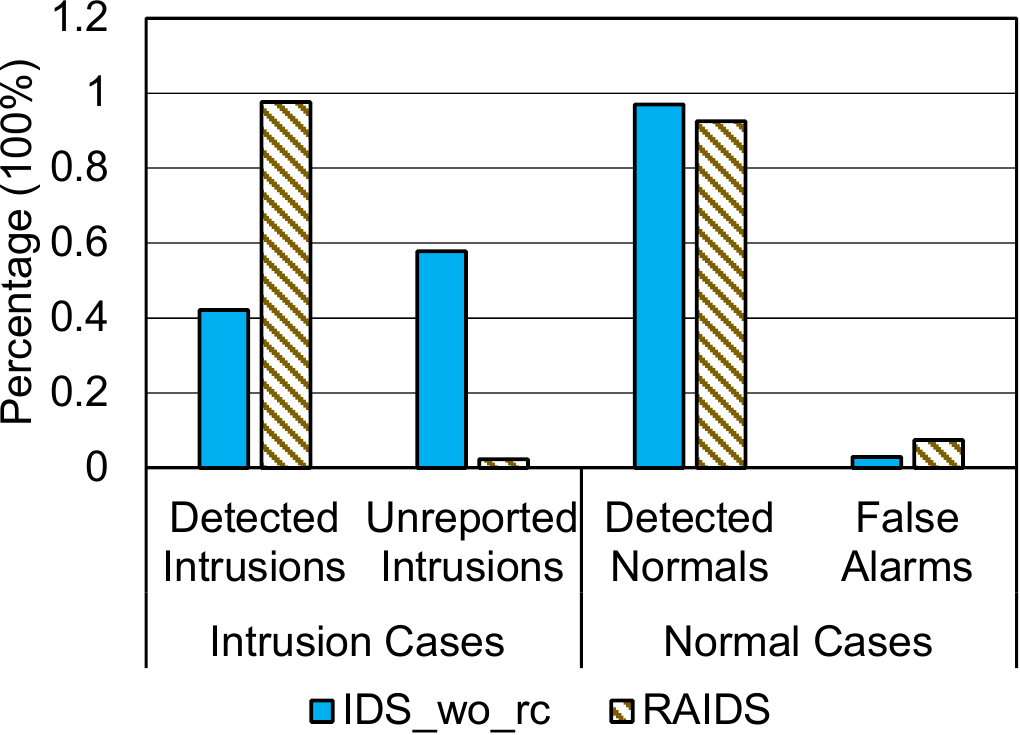}~\label{fig:false-random-udacitysim}}
	\subfigure[Udacity\_sim under Directed Intrusion ]{\includegraphics[width=0.485\columnwidth]{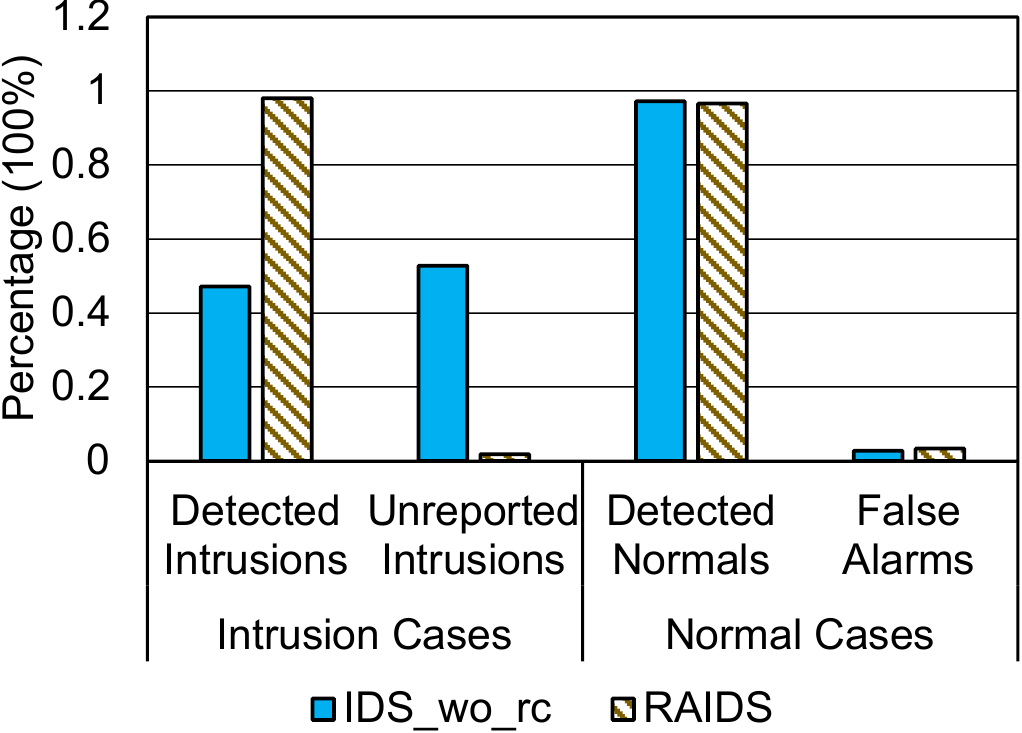}~\label{fig:false-directed-udacitysim}}
	\subfigure[Apollo under Abrupt Intrusion]{\includegraphics[width=0.485\columnwidth]{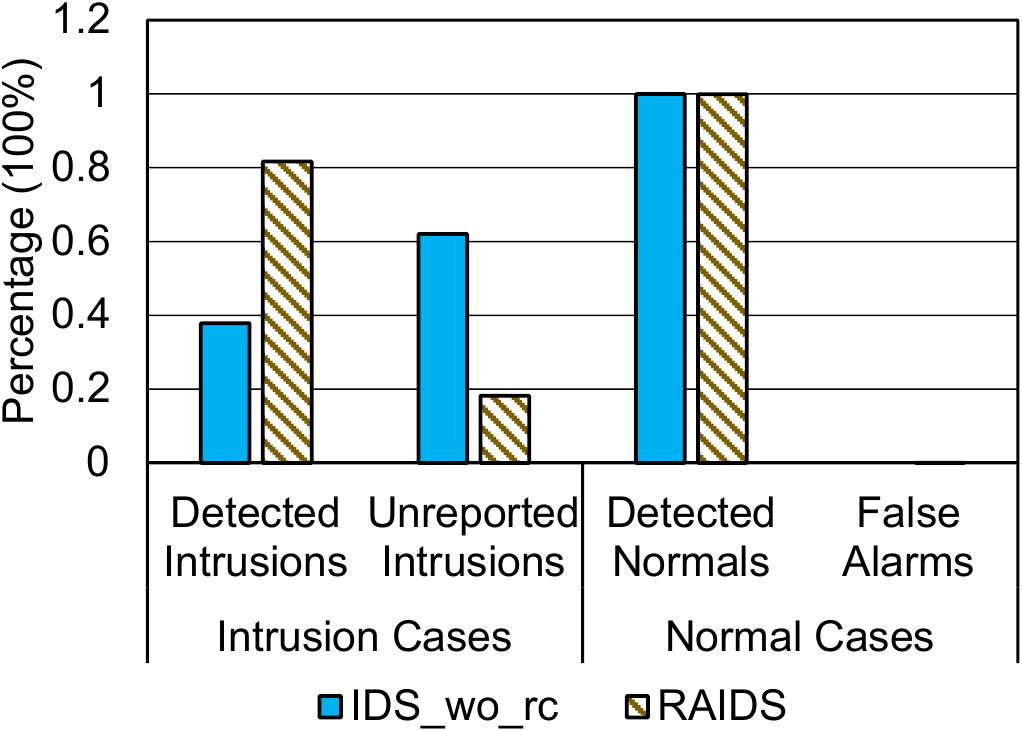}~\label{fig:false-random-apollo}}	
	\subfigure[Apollo under Directed Intrusion ]{\includegraphics[width=0.485\columnwidth]{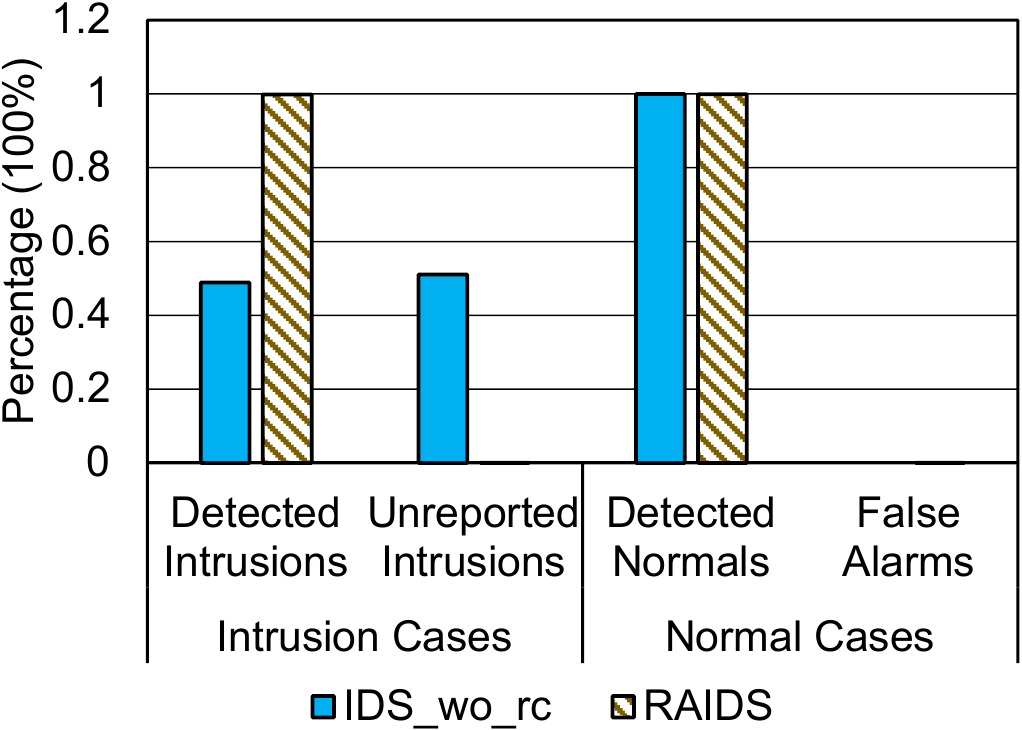}~\label{fig:false-directed-apollo}}
	\subfigure[Chen\_2017 under Abrupt Intrusion]{\includegraphics[width=0.485\columnwidth]{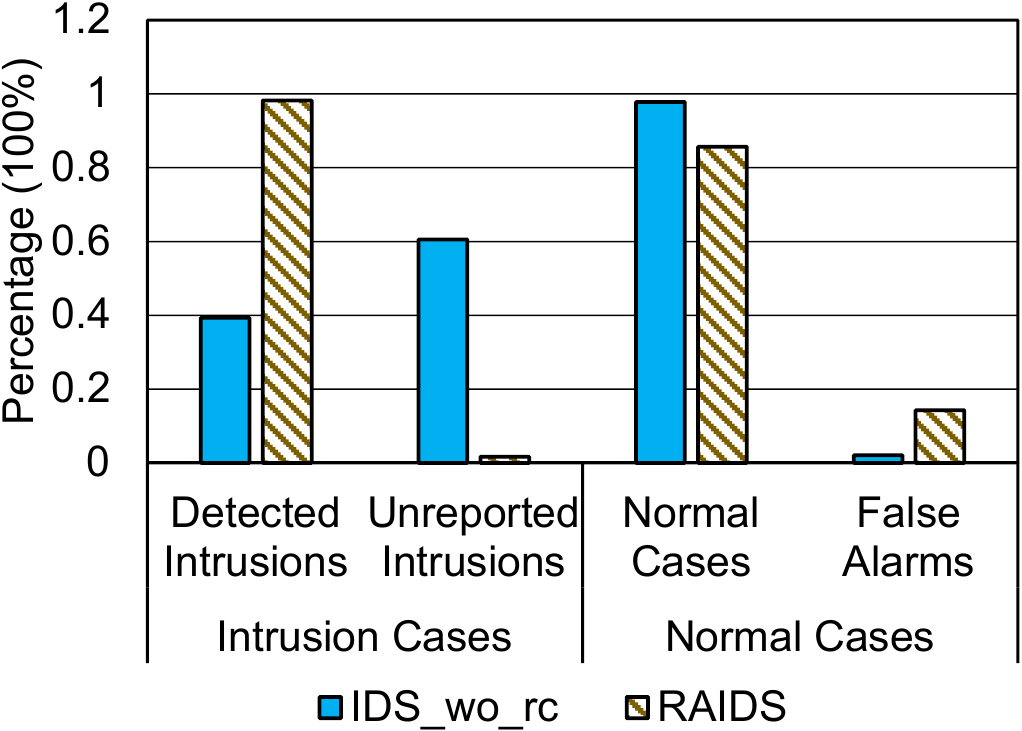}~\label{fig:false-random-chen2017}}
	\subfigure[Chen\_2017 under Directed Intrusion]{\includegraphics[width=0.485\columnwidth]{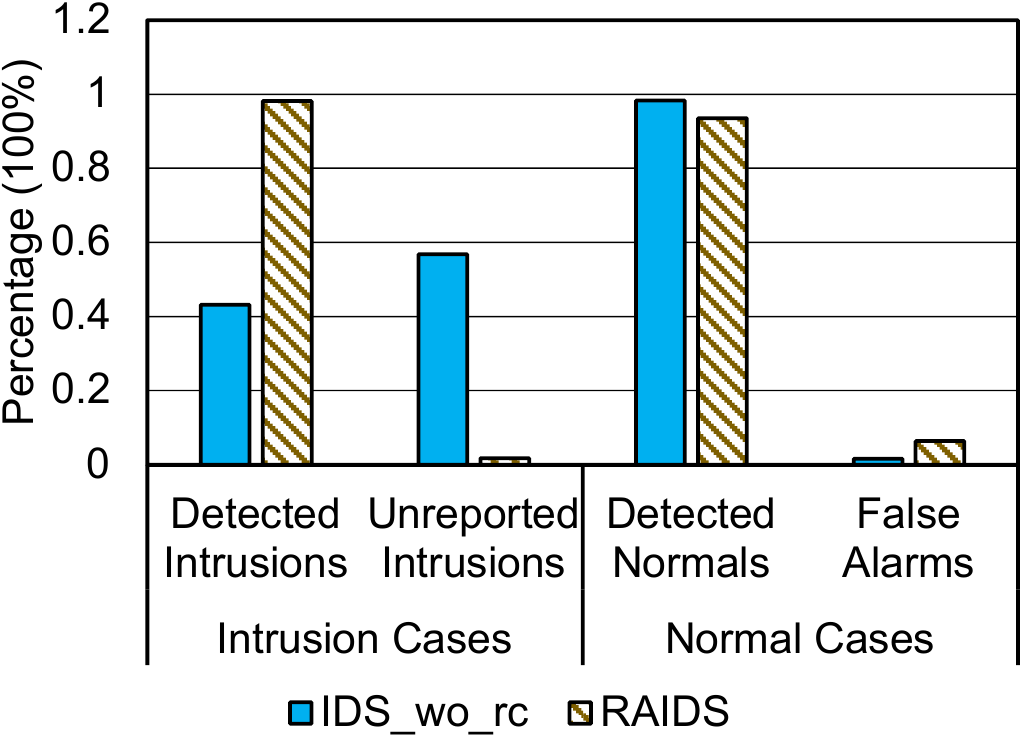}~\label{fig:false-directed-chen2017}}
	\subfigure[Chen\_2018 under Abrupt Intrusion]{\includegraphics[width=0.485\columnwidth]{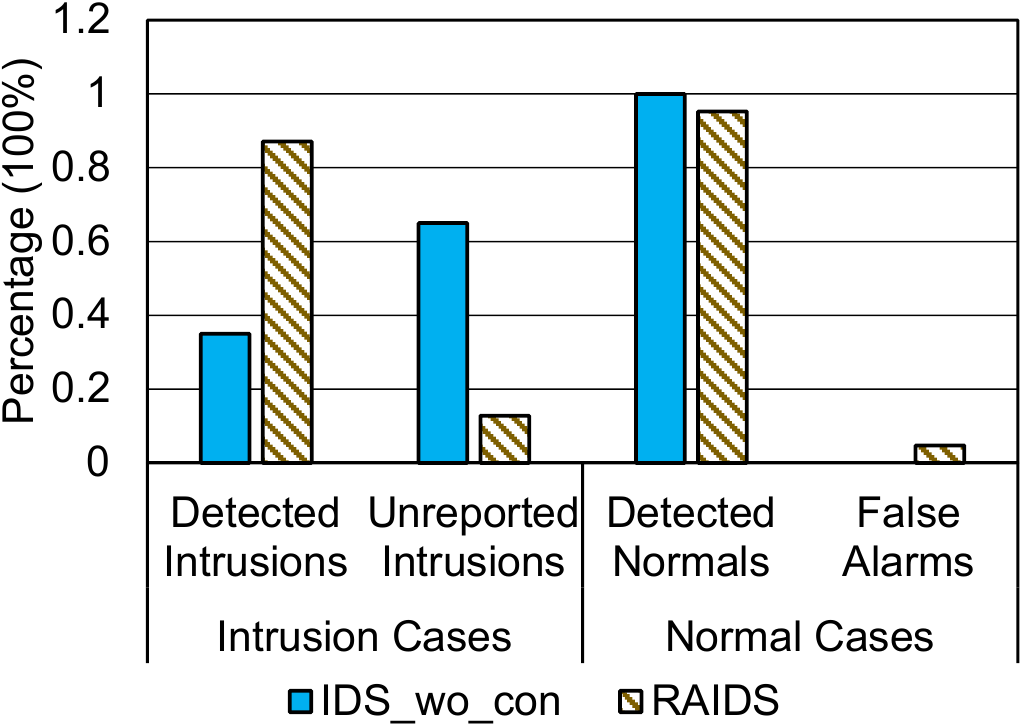}~\label{fig:false-random-chen2018}}
	\subfigure[Chen\_2018 under Directed Intrusion]{\includegraphics[width=0.485\columnwidth]{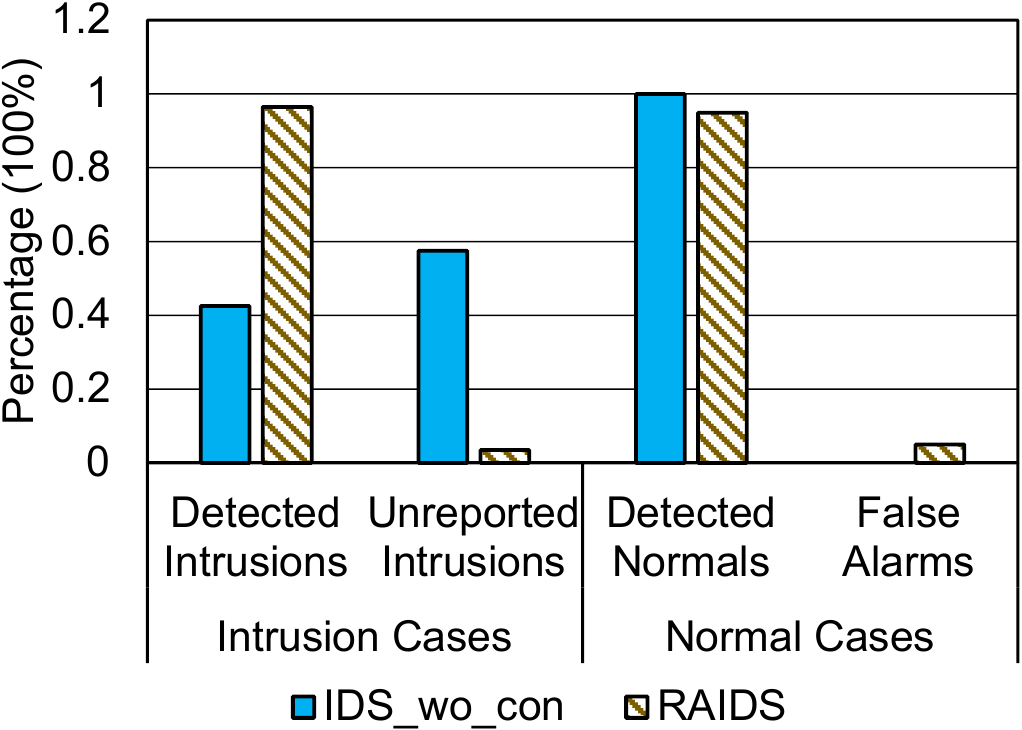}~\label{fig:false-directed-chen2018}}
	\subfigure[Comma.ai under Abrupt Intrusion]{\includegraphics[width=0.485\columnwidth]{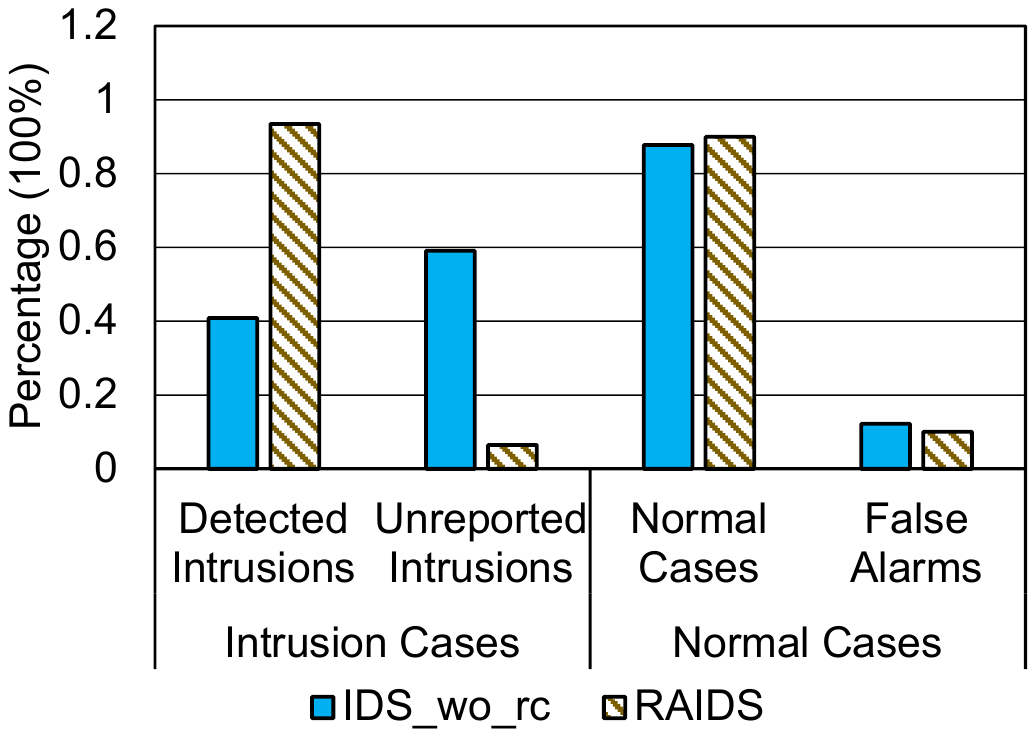}~\label{fig:false-random-commaai}}		
	\subfigure[Comma.ai under Directed Intrusion]{\includegraphics[width=0.485\columnwidth]{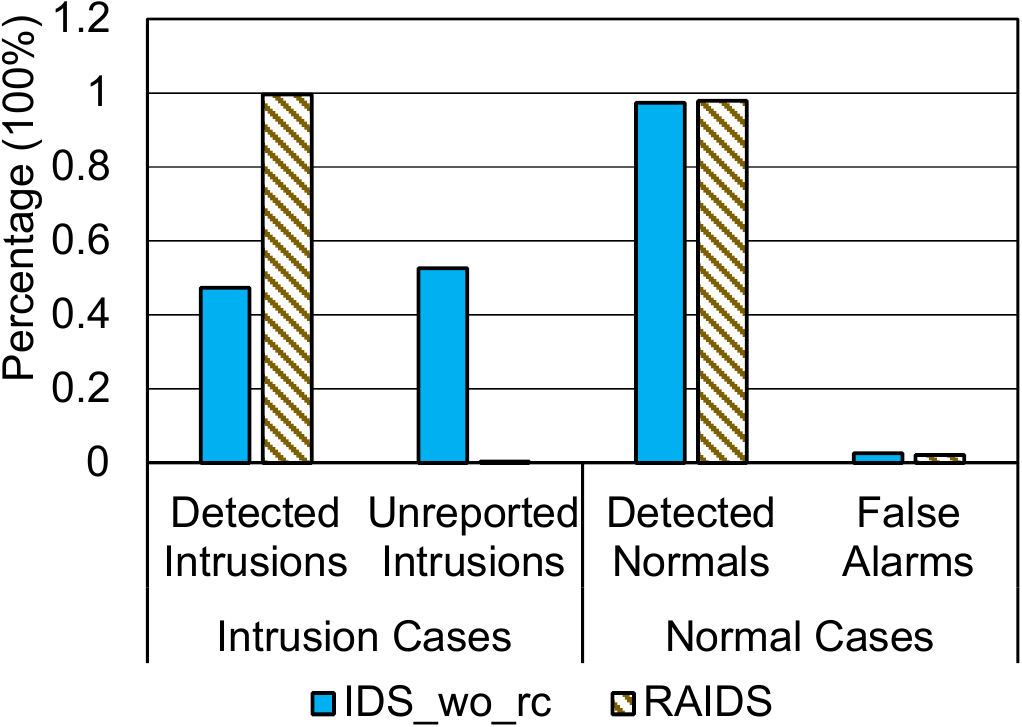}~\label{fig:false-directed-commaai}}			
	\vspace{-2ex}		
	\caption{The Percentages of Unreported Intrusions and False Alarms for Two IDSs under Abrupt and Directed Intrusions}\label{fig:false}
	\vspace{-2ex}	
\end{figure*}


\subsection{Detection Accuracy}

\textbf{Detection Accuracy}\hspace{2ex}
\autoref{fig:accuracy-summary} summarizes the detection accuracies of 
RAIDS and IDS\_wo\_rc with six datasets under abrupt and directed intrusions.
Two observations
can be obtained from~\autoref{fig:accuracy-summary}.
An apparent one is that, RAIDS consistently achieves high detection accuracies 
across different datasets under both abrupt and directed intrusions.
In particular, the highest accuracy for RAIDS is 99.9\% with Apollo
under directed intrusion while its lowest accuracy is 89.5\% 
with Comma.ai under abrupt intrusion. Comparatively,
 IDS\_wo\_rc yields lower detection accuracies. IDS\_wo\_rc's 
highest accuracy is 84.5\% with Apollo under directed intrusion 
while its lowest accuracy is 71.8\% with Comma.ai under 
abrupt intrusion. The significant gap between RAIDS's and IDS\_wo\_rc's 
accuracies confirms the high effectiveness of RAIDS. RAIDS differs 
from IDS\_wo\_rc in that RAIDS leverages the road context for intrusion
detection. IDS\_wo\_rc solely relies on the data of historical CAN frames
to apprehend the newly-arrived CAN frame. 
As shown in Figure~\ref{fig:speed}, 
the runtime volatile curve of steering angle 
alone is difficult to be modeled, 
unless it is associated with corresponding road context, like what RAIDS 
does. So the model built by IDS\_wo\_rc
lacks reliability. RAIDS, on the other hand,
extracts a feature vector
of road context from each image and involves the feature vector
for validating corresponding CAN frames. RAIDS thus establishes a sound model
that maps a specific road context, like a road bend shown in 
Figure~\ref{fig:bend}, to CAN frames. In summary,
if adversaries put frames with abnormal data on the in-vehicle CAN bus,
the unreliable model of IDS\_wo\_rc is ineffective in  
identifying the anomaly; however, on account of 
the involvement of road context, RAIDS has a high likelihood of
detecting the intrusion.

The second observation obtained from~\autoref{fig:accuracy-summary} is that
the detection accuracy under directed intrusion is consistently higher
than that under abrupt intrusion, especially for RAIDS.
For example, with Udacity, Apollo and Comma.ai datasets, the accuracy
of RAIDS under directed intrusion is 4.0\%, 5.4\% and 7.2\% higher  
than that under abrupt intrusion, respectively. As mentioned in Section~\ref{sec:attack},
directed intrusion should be more hazardous than abrupt intrusion because
the former intends to incur a sudden change at a specific occasion
onto the in-vehicle communications. 
Such a sudden change, whereas, brings in more significant violation
to ongoing road context, which exactly matches the capability of RAIDS
and can be easily captured.
This explains why RAIDS yields higher detection accuracy under 
directed intrusion.
On the other hand, as shown in Figure~\ref{fig:angle}, 
there exist dramatic increase and decrease of steering angle at runtime in reality. 
Consequently, IDS\_wo\_rc does not raise much difference in accuracy, 
i.e., at most 5.4\% with Apollo.

\begin{figure*}[t]
	\centering
	\subfigure[Abrupt Intrusion]{\includegraphics[width=0.99\columnwidth]{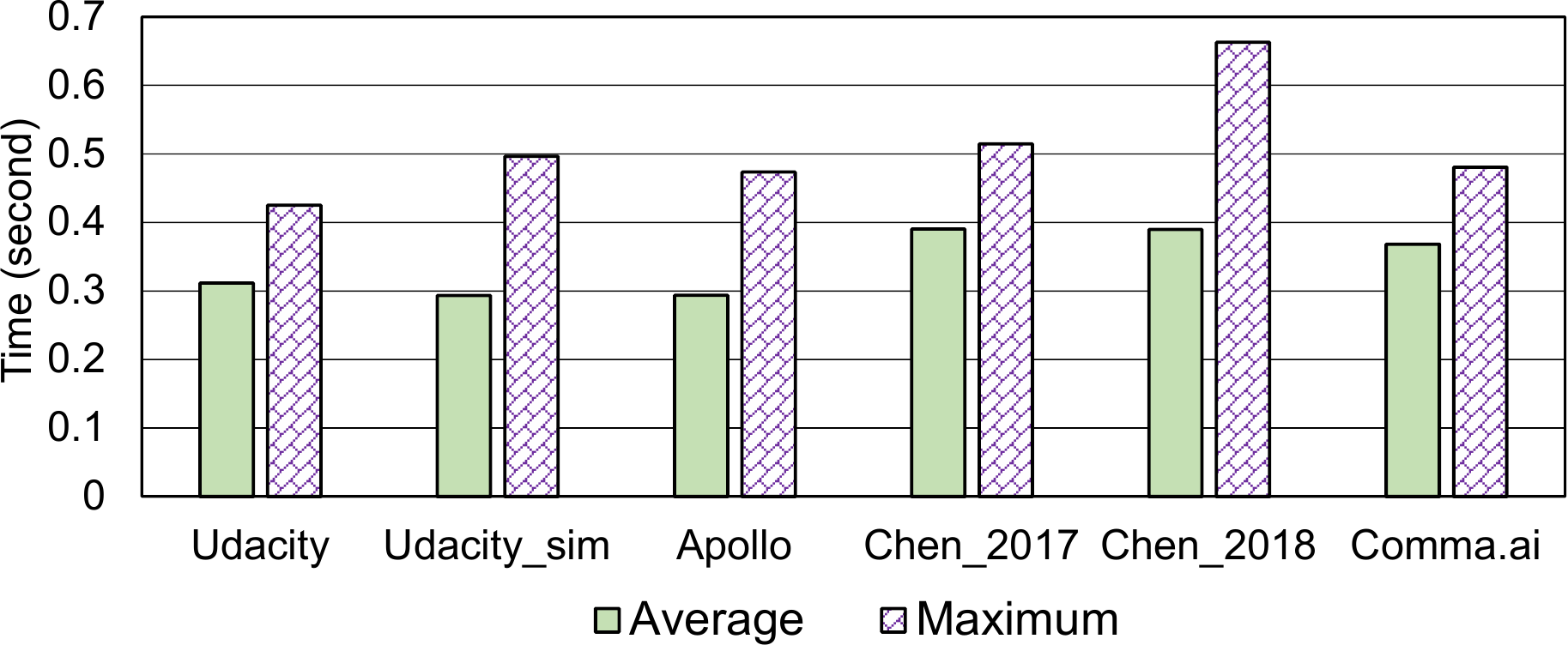}~\label{fig:random-time}}
	\hfill	
	\subfigure[Directed Intrusion]{\includegraphics[width=0.99\columnwidth]{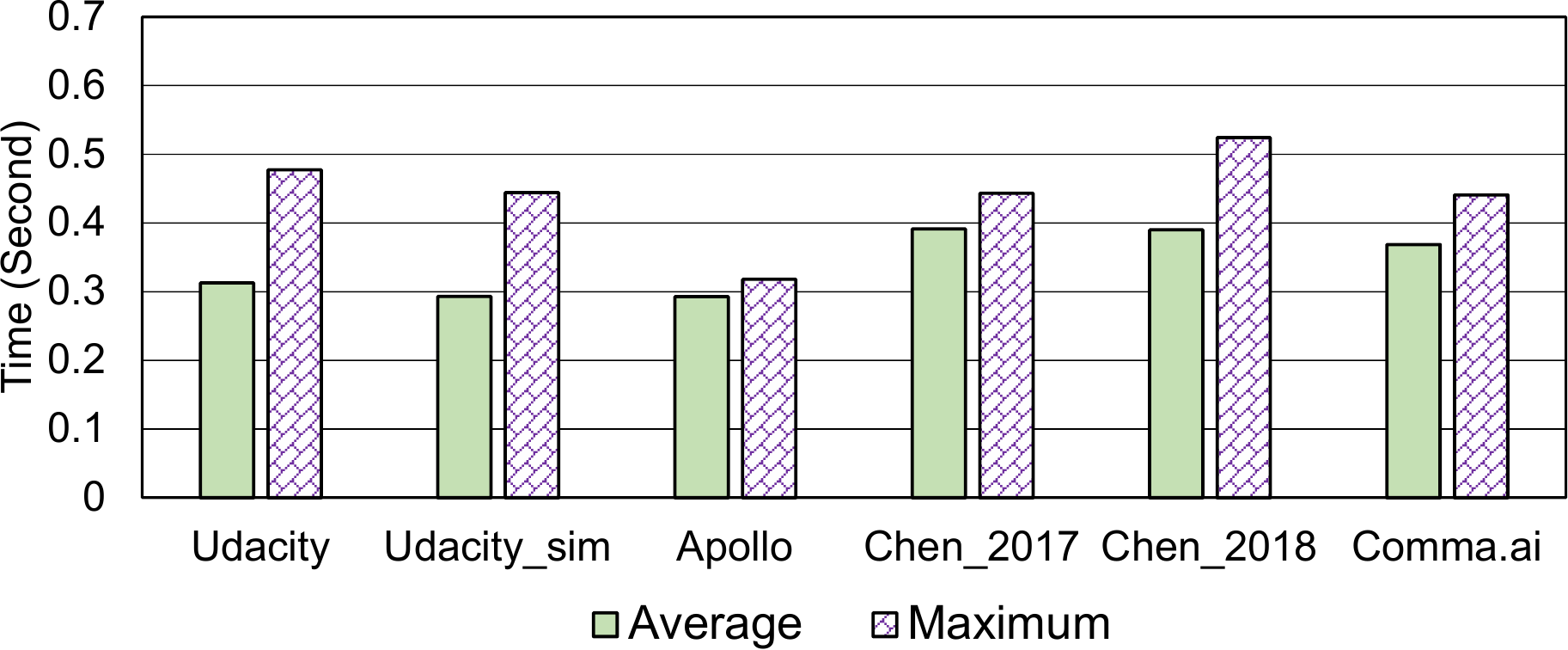}~\label{fig:directed-time}}
	\vspace{-1ex}		
	\caption{The Average and Maximum Response Time of RAIDS under Two Intrusions}\label{fig:time-summary}
	\vspace{-2ex}			
\end{figure*}

\textbf{Unreported Intrusions}\hspace{2ex}
We also record the percentages of detected and unreported intrusions as well
as detected normals and false alarms for six datasets under two types of intrusions.
These results help us gain a deeper understanding of the accuracies of 
RAIDS and IDS\_wo\_rc.
They are detailed in~\autoref{fig:false}. 
Let us first focus on the percentages of detected intrusion cases as the percentage
of false alarms are generally low.  
An obvious observations is
that, in all 16 diagrams, RAIDS detects most of the intrusion cases while 
IDS\_wo\_rc even cannot detect half of them. For example, 
in Figure~\ref{fig:false-random-chen2018}, the percentages of detected 
and unreported intrusion cases are 87.2\% and 12.8\%, respectively, for RAIDS with Chen\_2018 under 
abrupt intrusion;
however, they are 35.0\% and 65.0\% for IDS\_wo\_rc, respectively.
In other words, without considering road context, IDS\_wo\_rc ignores
many intrusion cases. 
This in turn justifies the importance of road context in intrusion detection.
In addition, as mentioned, the accuracy of Apollo under directed intrusion is 99.9\%.
From Figure~\ref{fig:false-directed-apollo}, we can see that there is hardly 
unreported intrusion or false alarm. This reaffirms the highest accuracy achieved by 
RAIDS.

Second, let us make a comparison between abrupt and directed intrusions.
Take Chen\_2018 dataset for example again with Figure~\ref{fig:false-random-chen2018} and Figure~\ref{fig:false-directed-chen2018}.
The percentage of detected intrusion cases for RAIDS increases 
from 87.2\% under abrupt intrusion to 96.5\% under directed intrusion. 
Such an increase confirms that directed intrusion 
is likely to incur intrusion cases that are easier to be perceived.
Comparatively, the percentage of detected intrusion cases for IDS\_wo\_rc also jumps from
35.0\% under abrupt intrusion to 42.5\% under directed intrusion. Although
the difference is considerable (42.5\% $-$ 35.0\% $=$ 7.5\%), 
it is less than that of RAIDS (96.5\% $-$ 87.2\% $=$ 9.3\%).
These numbers agree with the second observation we have had with~\autoref{fig:accuracy-summary}, 
and explain why the accuracy of IDS\_wo\_rc does not increase as much as that of RAIDS 
from abrupt intrusion to directed intrusion.

\subsection{Response Time}

With Raspberry Pi,
we have measured the response time of RAIDS in detecting intrusions for six datasets.
The two diagrams in 
~\autoref{fig:time-summary} capture the average and maximum
 response time of processing all records of images and corresponding CAN frames
 under abrupt and directed intrusions, respectively.
The results shown in~\autoref{fig:time-summary} 
state that none of the average response time is
greater than 0.40 second. 
More important, the maximum response time, which meas the worst-case response time,
is mostly no greater than 0.52 second, except with Chen\_2018 for which 
RAIDS cost 0.66 second under abrupt intrusion.
Concretely,
the short response time justifies the efficiency of RAIDS.

The average response time does not deviate much from the maximum response time
because each record contains almost the same quantity of data, i.e., image and CAN frames,
for RAIDS to handle. The marginal deviation is mainly 
caused by other running programs and system
scheduling in a real embedded system.
We note that we have used an economical Raspberry Pi for testing.
The computational resources
of Raspberry Pi includes   
an inexpensive 1.4GHz ARM CPU and 1GB DRAM.
With regard to employing RAIDS in real-world autonomous cars, a 
more powerful embedded system with 
high-end CPU and large RAM space could be 
leveraged to reduce the response time, which surely improves efficiency and 
applicability of RAIDS.

\subsection{The Impact of Daytime and Night}

\begin{figure}[t]
	\centering
	\subfigure[Detection Accuracy]{\includegraphics[width=0.47\columnwidth]{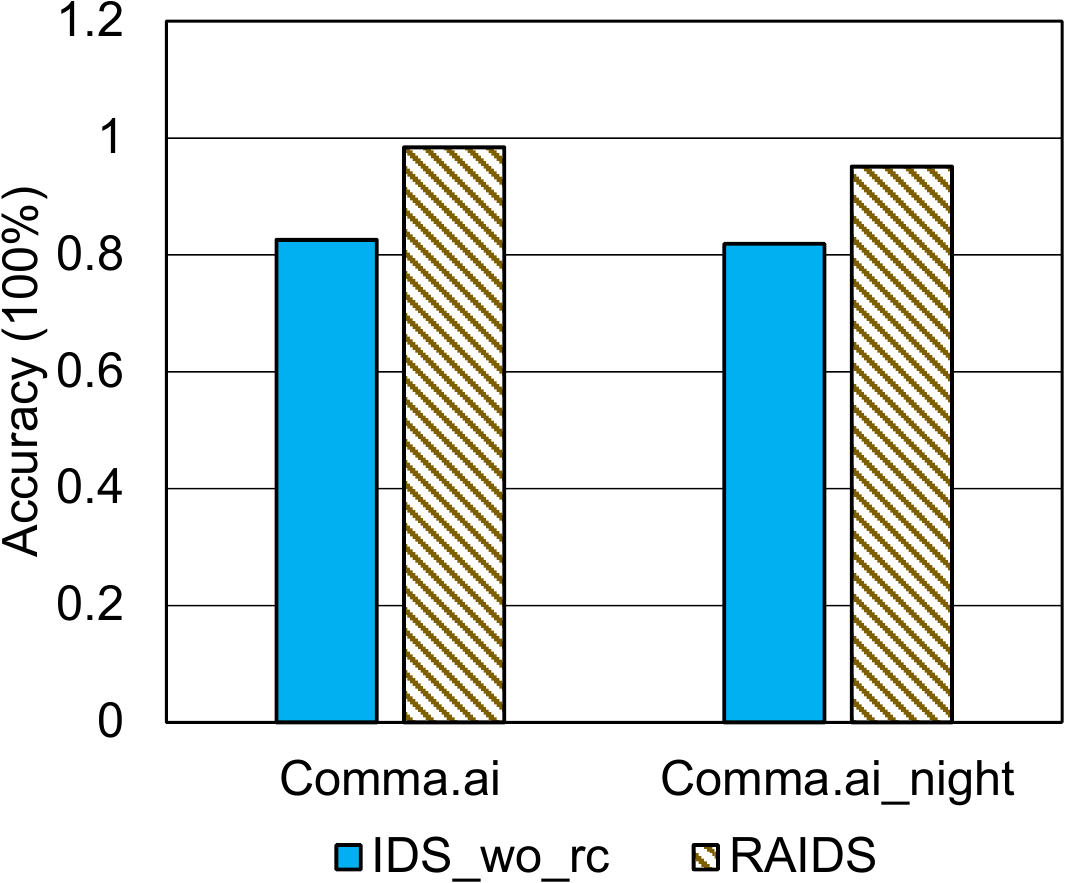}~\label{fig:night-accuracy}}
	\hfill
	\subfigure[Unreported Intrusions and False Alarms with Comma.ai\_night]{\includegraphics[width=0.47\columnwidth]{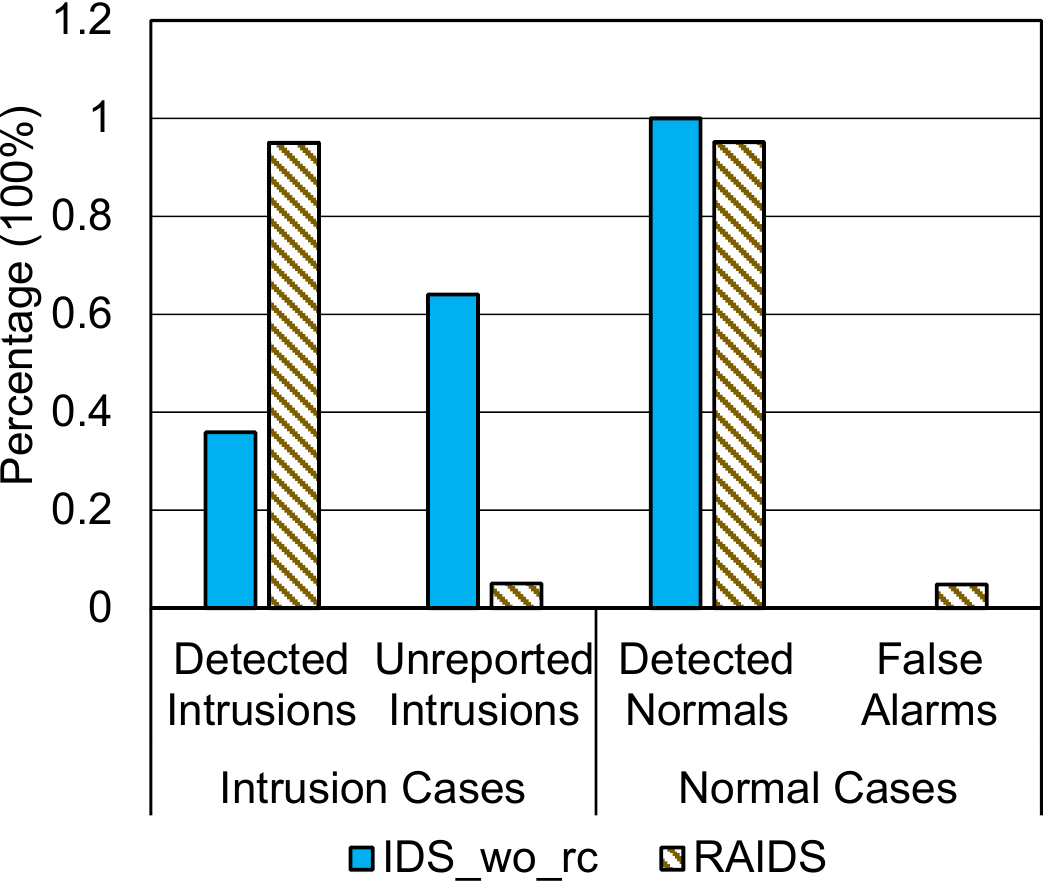}~\label{fig:night-false}}
	\caption{A Comparison between Daytime and Nighttime Datasets for RAIDS}\label{fig:day-night}
	\vspace{-3ex}		
\end{figure}

It is non-trivial to extract meaningful features from nighttime images
due to the generally low visibility of road conditions~\cite{car:night:IEEE-2017}.
In fact, Uber's fatal accident happened at night 
when a pedestrian was crossing a road~\cite{car:Uber-accident:2018}.
Comma.ai provides a sub-dataset with nighttime images and CAN frames (referred to as
Comma.ai\_night).
We have done abrupt and directed intrusions with it. 
Because of space limitation, we present the results under directed intrusion.
Figure~\ref{fig:night-accuracy}
captures the comparison of IDS\_wo\_rc's and RAIDS's accuracies 
between daytime Comma.ai and nighttime Comma.ai\_night.
In fact, the accuracy of IDS\_wo\_rc does not fluctuate much since it 
is oblivious of the change of day and night. Nevertheless,
due to the weaker perception of road context at night,
the accuracy of RAIDS drops by 3.3\%. 

Let us do a comparison between Figure~\ref{fig:false-directed-commaai} for Comma.ai and 
Figure~\ref{fig:night-false} for Comma.ai\_night. In particular,
from Figure~\ref{fig:false-directed-commaai} to Figure~\ref{fig:night-false},
the percentages of unreported intrusions and false alarms for RAIDS increase
by 4.6\% and 2.7\%, respectively. In other words, a bit
more intrusion and normal cases have been wrongly deemed to be anomalous 
with regard to the relatively obscure nighttime road contexts. 
This exposes the reason for the accuracy drop of RAIDS.

%% file: conclusion.tex
\section{Conclusion}

Security is always critical for traffic vehicles.
In this paper, we investigate how to effectively detect 
intrusions to the in-vehicle communications of autonomous car. 
In an unmanned autonomous car, a self-driving model 
reads sensory information reflecting dynamic road contexts and 
generates control signals 
that are eventually transformed into frames transmitted on the CAN bus.
The ever-changing control signals at runtime   
require the involvement of road contexts in detecting intrusions 
for autonomous car. We accordingly develop RAIDS. RAIDS
extracts and abstracts road contexts 
from sensory information  into a feature vector.
It then leverages such an expressive feature vector of road context to 
assert the genuineness of 
observed CAN frames.
We have built a prototype for RAIDS through lightweight neural networks,
and evaluated it in an embedded computing system with extensive datasets. 
Experimental results confirm that RAIDS achieves up to 99.9\% accuracy with 
short response time.